\documentclass[10pt]{article}
\usepackage[explicit]{titlesec}
\usepackage{hyphenat}
\usepackage{ragged2e}
\RaggedRight

\usepackage{rotating}  
\usepackage{theorem}   
\usepackage{bm}  
\usepackage{amsmath,amsfonts,amssymb}
\usepackage{booktabs}  
\usepackage{pdfsync}   
\usepackage{graphicx}
\usepackage{latexsym}
\usepackage{amsmath}
\usepackage{amssymb}
\usepackage{fancyvrb}
\usepackage{subfigure}
\usepackage{colortbl}
\usepackage{enumerate}
\usepackage{pifont}
\usepackage{stmaryrd}
\usepackage{textcomp}
\usepackage{fncylab}
\usepackage{multirow}
\usepackage{paralist}
\usepackage{wrapfig}
\usepackage{longtable}
\usepackage[]{algorithm2e}
\usepackage[table]{xcolor}
\usepackage{lscape}
\usepackage{lipsum}

\usepackage{garamondx}
\usepackage[T1]{fontenc}
\usepackage{amsmath}
\usepackage{graphicx}
\usepackage{xcolor}

\usepackage{geometry}
\usepackage{fancyhdr}
\usepackage[labelfont={footnotesize,bf} , textfont=footnotesize]{caption}
\makeatletter
\renewcommand\tagform@[1]{\maketag@@@ {\ignorespaces {\footnotesize{\textbf{Equation}}} #1.\unskip \@@italiccorr }}
\makeatother
\titleformat{\section}
  {\normalfont}{\thesection}{1em}{\MakeUppercase{\textbf{#1}}}
\titlespacing\section{0pt}{0pt}{-10pt}
\titleformat{\subsection}
  {\normalfont}{\thesubsection}{1em}{\textit{#1}}
\titlespacing\subsection{0pt}{0pt}{-8pt}

\makeatletter
\newcommand\sixteen{\@setfontsize\sixteen{17pt}{6}}
\renewcommand{\maketitle}{\bgroup\setlength{\parindent}{0pt}
\begin{flushleft}
\sixteen\bfseries \@title
\medskip
\end{flushleft}
\textit{\@author}
\egroup}
\makeatother

\usepackage[sort&compress]{natbib}
\makeatletter
\renewcommand\@biblabel[1]{\textbf{#1.}\hfill}
\makeatother

\bibpunct{}{}{,~}{s}{,}{,}
\title{Towards Knowledge-based Mining of Mental Disorder Patterns from Textual Data}

\author{
Maryam Shahabikargar*$^{a}$ \\ \medskip
$^{a}$Macquarie University, Sydney, Australia \\  \medskip
maryam.shahabi-kargar@hdr.mq.edu.au
}

\begin{document}

\vspace*{.01 in}
\maketitle
\vspace{.12 in}

\section*{abstract}

Mental health disorders may cause severe consequences on all the countries' economies and health. For example, the impacts of the COVID-19 pandemic, such as isolation and travel ban, can make us feel depressed. Identifying early signs of mental health disorders is vital. For example, depression may increase an individual's risk of suicide. The state-of-the-art research in identifying mental disorder patterns from textual data, uses hand-labelled training sets, especially when a domain expert's knowledge is required to analyse various symptoms. This task could be time-consuming and expensive. To address this challenge, in this paper, we study and analyse the various clinical and non-clinical approaches to identifying mental health disorders. We leverage the domain knowledge and expertise in cognitive science to build a domain-specific Knowledge Base (KB) for the mental health disorder concepts and patterns. We present a weaker form of supervision by facilitating the generating of training data from a domain-specific Knowledge Base (KB). We adopt a typical scenario for analysing social media to identify major depressive disorder symptoms from the textual content generated by social users. We use this scenario to evaluate how our knowledge-based approach significantly improves the quality of results.

\section*{keywords}
Cognitive Science, Knowledge Base, Machine learning; Business Process Analytics

\vspace{.12 in}


\section{introduction}

We begin this Section with an overview of the research problem and challenges in identifying and understanding mental health disorder patterns from textual data.
We present our contributions and discuss how the proposed method may facilitate acquiring insight into the mental status of individuals who may be suffering from mental disorders in general and depression in particular.
Finally, we present the structure of this paper.

\subsection{Overview and Research Problem}

The mental health of individuals and communities is a pressing challenge in the world, nowadays.
Since COVID-19\footnote{https://en.wikipedia.org/wiki/COVID-19} pandemic outbreak
in 2019, most governments have been preoccupied with handling and combating the epidemic. The prevalence of the Covid-19 has been significantly controlled and lowered as a result of the global success in vaccine development and mass inoculation. But now, for most governments, a key concern is harnessing and dealing with the impacts of years of virus exposure, including the associated psychological and economic problems.
COVID-19 has impacted our lifestyles and workplaces. These changes can cause us to feel frustrated, stressed, and anxious,
which may seriously affect
our mental health. Based on a recent study, after being diagnosed with Covid, roughly one out of every five people develops a mental disorder~\cite{taquet2021bidirectional,taheri2021novel}. Hence, governments are trying to guide
families and business owners
to deal with these devastating effects and improve the situation.
%
On the other hand, as businesses reopen, it is important for organisations to provide a mentally healthy workplace\footnote{https://covid19.swa.gov.au/collection/covid-19-resource-kit}.

Prior to any support, mental disorders
need to be diagnosed.
On the other hand, due to their complexity, identifying mental disorder symptoms~(e.g., depression symptoms) and their patterns
could be a challenging task.
Hence, it is necessary to
identify mental health issues accurately and facilitate their treatment.
As a critical mental health issue, depression is
one of the leading causes of disability worldwide.
It plays
an essential role
in the overall global disease burden~\cite{de2013predicting, lemoult2019depression} and could turn into a drastic health condition~\footnote{https://www.who.int/news-room/fact-sheets/detail/depression}. Depression is a leading cause of disability, with 5\% of adults and 5.7\% of 60-year-old and above people suffering from it.
Data from the United States and Australia show
elevated rates of depression and anxiety throughout the epidemic. It is estimated that during the outbreak, depression level~(i.e., 25\%) is seven times higher than pre-pandemic levels worldwide~(i.e., less than 4\%)\footnote{https://www.forbes.com/sites/debgordon/2021/05/28/mental-health-got-worse-during-covid-19-especially-for-women-new-survey-shows/}. Consequently, due to its importance, we focus on depression identification in this research.

There are various clinical and non-clinical approaches available to identify depression symptoms.
Using questionnaires and interviews are two main clinical approaches for depression identification.
On the other hand, analysing medical data~(e.g., EEG and fMRI images) and vocal, video, and textual data are different approaches to identifying and predicting depression.
In addition, there are very few recent studies
that proposed
knowledge-based approaches to identify behavioural and mental disorders such as depression.
%

To extend the state-of-the-art in this line of work, in this paper,
%
we propose a domain-specific
Knowledge Base (KB)
%
to organise complex structured and unstructured clinical knowledge about symptoms, risk factors
and supportive symptoms
that are effective in identifying mental disorders.
%
We introduce a pipeline that leverages the KB knowledge to facilitate the identification of mental disorder patterns in textual data.
We evaluate our approach with a dataset,
generated from social media activities,
and highlight how the proposed framework can help analysts in the e-Safety\footnote{https://www.esafety.gov.au/} community to gain insight into the mental status of potential individuals
suffering from depression-related symptoms.

\subsection{Contributions}

%
The state-of-the-art research in identifying mental disorder patterns from textual data, uses hand-labelled training sets, especially when a domain expert's knowledge is required to analyse various symptoms. This task could be time-consuming and expensive. In this paper, we present a weaker form of supervision by facilitating the generating of training data from a domain-specific Knowledge Base (KB).
%
The knowledge in the KB is gathered from both cognitive and psychological sciences as well as previous best practices in the field.
This KB contains a set of concepts organised into a taxonomy, instances for each concept, and relationships among them.
The KB then will be used for developing a weakly supervised~\cite{ratner2019weak} classifier for mental disorder identification.
We propose a method to link the concepts and instances in the KB to the features extracted from textual data.


We discuss a motivating scenario
in which leveraging a
domain specific KB for depressive patterns
could enable e-safety community to
monitor the trend for the emergence of depression-related symptoms, during special periods of time such as the Covid-19 period.
This paper makes the following contributions:

    %
    %

\begin{itemize}
    \item We leverage the domain knowledge and expertise in cognitive science to build a domain-specific Knowledge Base (KB) for the mental health disorder concepts and patterns. This KB contains a set of concepts organised into a taxonomy, instances for each concept, and relationships among them.

    \item We present a weakly supervised learning approach
       by facilitating the generating of training data from a domain-specific Knowledge Base (KB).

    \item We adopt a typical scenario for analysing social media to identify major depressive disorder symptoms from the textual content generated by social users. We use this scenario to evaluate how our knowledge-based approach significantly improves the quality of results.
\end{itemize}


\section{Background and State-of-the-Art}
\label{Background_Section}

The fundamental part
of each scientific research is to study previous related works.
%
In this Section,
we review
the literature with respect to our research problem, namely mental disorder identification, focusing on depression,
a mental health condition marked by a consistently depressed mood or a loss of interest in things, resulting in severe impairment in everyday life.

There are numerous
studies
each address this problem from a
different point of views.
%
In this Section,
we study and analyse the recent work in mental disorders and cognitive analytics.
We discuss
Cognitive science and cognitive analytics and the way they are connected to mental disorders
In addition, some cognition-related mental disorders such as depression are explained from a psychological and cognitive analytics perspective.
Finally, we study and analyse the related work in clinical and non-clinical approaches for identifying depression symptoms.

\subsection{Mental Disorders and Cognitive Analytics}

Mental health issues are proved to play a significant role in reaching
worldwide
developmental objectives. Mental disorders have
risen
notably during the past decades. They caused severe consequences on
the economy
and health of all the countries~\cite{Whomentalhealth}. Mental disorders have a significant role in
every aspect
of people's life. It affects the way people behave, act and work, while having
a significant
influence on people's physical health and their personal and social relationships.
Mental health issues have drastically affected
most countries.
For example, nearly one-half of Australian
adults, including
7.3 million individuals,
would experience mental-related
matters
at some point in life. These issues cause Australia's economy an approximate cost of up to 220 billion
dollars yearly.

Mental health issues could negatively impact anyone at any life-stage as well as people around~them~\cite{ausgovmentalhealth}.
There are several factors such as genetic, drug and alcohol abuse, early life circumstances, trauma, stress and personality factors contributing to mental disorders \footnote{https://www.healthdirect.gov.au/mental-illness}. In addition, a recent study found that the recent global pandemic~(i.e., Covid-19) has caused nearly one in every five patients
to develop
a mental disorder, after being diagnosed with Covid. Also, those with previous mental conditions are 65\% percent more likely to be diagnosed with Covid-19, even when other risk factors are considered~\cite{taquet2021bidirectional}.
Despite extremely varied Covid-19 prevalence and
fatality levels,
research from the United States and Australia demonstrate heightened rates of depression and anxiety throughout the epidemic. The
approximated
level of depression in the outbreak~(i.e., 25\%) is seven times greater than
the ratio before it worldwide
(i.e., less
than 4 percent)~\cite{mentalhealthduringcovid}. Hence, the importance of addressing mental health and facilitating their treatment has become much more prominent.

Mental disorders could be classified
into
several categories, such as `Anxiety Disorders', 'Depressive Disorders', `Neurocognitive Disorders', `Paraphilic Disorders' and `Bipolar and Related Disorders'. Each of these categories consists of different instances of that category. For example,
some cases
of depressive disorders are `Premenstrual Dysphoric Disorder', `Disruptive Mood Dysregulation Disorder', `Premenstrual Dysphoric Disorder' and `Substance/Medication-Induced Depressive Disorder' and `Major Depressive Disorder~(MDD)'. Throughout this study, mainly, we refer to MDD as depression.
Also, the category of `Anxiety Disorders' contains some instances such as `panic disorder', `social anxiety disorder', `agoraphobia' and `generalized anxiety disorder'.

Mental disorders are originated from psychological processes impairment. Psychological processes
consist of
cognitive
processes (i.e., the main factor),
Individual experiences, behavioural, social and biological factors~\cite{kinderman2005psychological}.
a key
element that contributes
to mental disorders such as `Bipolar and Related Disorder', `Sleep Disorders' and `Depressive disorders' is cognitive processing changes that remarkably affect the individual's capacity to function.~\cite{edition2013diagnostic, hammar2009cognitive}. Cognitive Science, referring to the scientific study of mind and brain~\cite{friedenberg2021cognitive, nadel2003cognitive}, is mainly
related to understanding the nature of cognitive processes such as problem-solving, language, perception, reasoning, attention and motor control~\cite{cogsciyaleuni}.

More acceptable
recognition of the relationships between cognitive processing and mental disorders is necessary for treatment interventions~\cite{o2008personality}. On the other hand, one of the main advantages of data science and machine learning~(ML) techniques are finding various relationships among variables. Hence, using data analytics and cognitive analytics helps researchers
see the relationships
between different variables related to cognitive processing and mental disorders. Consequently, developing ML models for identifying mental and behavioural disorders patterns would be enabled~\cite{barukh2021cognitive}.

Cognitive analytics means data processing aiming at understanding varied, intricate, heterogeneous and qualitative data~\cite{handfield2019emerging}. It could be considered as doing analytics,
originating
from a human-like intelligence. Cognitive analytics refers to the advanced approaches integrating the cognitive science knowledge~(e.g., the logic behind human decision-making, thought patterns and cognitive processes) with data analytics techniques
(e.g., ML
and artificial intelligence and other technologies related to data science)~\cite{handfield2019emerging}.

Cognitive analytics could help with recognising
the natural human
language and interactions~\cite{coganalytics}, through analysing both structured and unstructured data~\cite{handfield2019emerging} such as text, conversations, email, social posts,
images
and video data. It also allows cognitive applications to improve continuously, while could be used to help companies understand the emerging trends and behavioural
patterns
of their customers. Hence, they would be able to predict the probable future consequences and plan ahead for their objectives~\cite{coganalytics02}.

There are several studies related to applications of data and cognitive analysis in identifying and predicting mental and behavioural disorders~\cite{de2013predicting, majumder2017deep, rezvani2020linking}. Cyberbullying~\cite{rezvani2021towards}, radicalisation, and
suicidal-related
behaviours are examples of such applications. Social media is a valuable source of data for analytical purposes, making
preventing
and intervening actions in such disorders possible~\cite{beheshti2022social}.
In the following sections,
these three
applications are briefly explained. In addition, depression and anxiety are briefly described as two major mental disorders associated with cognitive impairments~\cite{ahern2017cognitive, chakrabarty2016cognitive}.

\subsubsection{Cyberbullying }

Cyberbullying is known as an online bullying matter. It is a kind of antisocial behaviour
that may include
aggression and intentional power abuse
in a cyber environment
(e.g., social media or internet-based platforms), leading to detrimental social, mental  and psychological effects. Sending
inappropriate
online messages and comments with adverse content could be indicative of cyberbullying actions~\cite{rezvani2020linking}.
Cyberbullying
gained considerable attention from researchers and governments because of the negative psychological issues and problems
that arise
from it~\cite{yudes2020predictive}.

The brain could be influenced by cyberbullying in various ways depending on the
age, and
gender~\cite{skilbred2020cyberbullying}. Cyberbullying appeared to be more and more common among adolescents\footnote{https://en.wikipedia.org/wiki/Cyberbullying}. In adolescence, there are significant changes
that may happen
in the brain's
developmental
processes, which
consequently cause
substantial changes
in
adolescents'
behaviour, cognition and emotion~\cite{blakemore2006development}. Hence, they could be more vulnerable in terms of being affected by adverse events and
experiences
(e.g., cyberbullying behaviours), which in turn cause them various emotional, mental and behavioural problems~\cite{mcloughlin2020cyberbullying}.
Changes in mental health status,
and decision-making processes,
and self
and emotional regulations are all examples of changes that could happen during adolescence. Cyberbullies have poor brain executive functions, low empathy and social cognition. Cyberbully victims are prone to suffering from mental disorders, specially suicide attempts~\cite{skilbred2020cyberbullying, mcloughlin2020cyberbullying}. Brain executive function refers to the capacity for controlling and coordinating thoughts and behaviour. Also, social cognition capacity encompasses self-awareness and the ability to understand other
people's
minds~\cite{blakemore2006development}.

Kinds of words~(e.g., self-focused words, hate speech, positive or negative emotion, death-related, anger), sentence sentiments and natural language that are used in the daily communications are reflections of individuals' real-life manners, thoughts, cognitive processes, behaviours, and mental health status~\cite{glenn2020can, serani2011living, al2018absolute, de2013predicting, tausczik2010psychological, schmidt2019survey,  troop2013expressive}. As the studies show, Cyberbullies tend to
use a high rate of profane and vulgar words, which
is demonstrated
in their
cyberbully activities in
an internet-based environment
such as social media~\cite{rezvani2020linking, tarmizi2020detecting}. Social media, as a virtual daily communication platform, is proved to be a rich source of different types of data, especially
textual data.
Social media enables researchers and
data scientists
analyse cognitive and behavioural processes and patterns related to the social media users~\cite{de2013predicting}.
As an example application of data science in finding cognitive and mental-related patterns,
Rezvani et~al.,
Proposed a Cyberbullying detection pipeline for social media data~(i.e., Twitter and Instagram).
It benefits
from several data analytics techniques such as `Natural Language Processing~(NLP)', `Neural Networks~(NN)', and `Long short-term memory~(LSTM)'\cite{rezvani2020linking}. Through their pipeline, different types of textual and contextual features, image-related features and metadata features are extracted with the aim of developing a cyberbullying identification model~\cite{tabebordbar2020feature}.

Textual features include
Parts of speech, named entities and sentiment
analysis, etc.
In this cyberbullying pipeline, TF-IDF and NLP methods
are used to extract
textual features. Besides, image-related features~(i.e., labels for images of social media posts) and 
metadata
are extracted.
Metadata
features mainly refer to behavioural information of the corresponding social media user.
Metadata features that are
related to social media
include
`Number of followers', `Number of followees', `Number of likes', `Popular categories', `Average reactions', `Average replies', `Frequent mentions'. They also used
Google standard profanity word list aiming at enriching the feature extraction process. Finally, they built an LSTM classifier along with Neural Network for context2vec embeddings that 
combine
features to recognise probable useful features~\cite{abu2021relational}.
Their proposed model had notable evaluation results for both Twitter and Instagram data. Twitter data analysis illustrated 0.85 accuracy, 0.87 Precision, 0.83 Recall, and 0.85 F-score. Similarly, the results of Instagram data analysis showed 0.86 accuracy, 0.87 Precision, 0.83 Recall, and 0.85 F-score. All these results are indicative of promising results originated from cognitive  and data analytics in identifying mental and behavioural disorders.

\subsubsection{Radicalization }

Radicalization is
known to be the mechanism of alteration in
individuals'
cognitive processing,
beliefs,
emotions and consequently their behaviours. These changes lead to legitimizing inter-group violence and the need for sacrifice in favour of one's own group~\cite{bott2009recruitment}. Cognitive opening refers to the moment when the ideology of thinking-radicalised as a cognitive process is shaped within a person. It happens when accepted beliefs of a potential individual who is experiencing political and socioeconomic discrimination are shaking, and he/she becomes susceptible to accept those ideology\cite{wiktorowicz2005radical}.
Some of the cognitive
factors that assist
the cognitive opening are feeling deprived, perceiving unfairness, injustice,
societal disconnectedness,
and societal disconnectedness;
whether as an individual or as a group member. Besides, feeling symbolic threats along
within-group
superiority are considered to be other effective cognitive factors. Feeling symbolic threats occur when in-group members believe out-group
individuals are acting in a way that is harmful to their in-group members,
consequently getting the belief that their group's moral and values are more important. All these factors
may help in feeding
violent attitudes~\cite{trip2019psychological}.

Influential cognitive and behavioural factors, which are causing individuals to experience mental health issues, could be analysed
by leveraging
predictive analysis. As an application of data science analysis along with cognitive science, Beheshti etal.~\cite{beheshti2021towards} proposed a pipeline to help analysts with exploring potential extremist activity patterns through social media data. With the aim of anlysing those factors, They introduced some new concepts such as
a cognitive graph,
entity, and relationship. The cognitive graph enables exploration and interpretation of significant cognitive and behavioural patterns existing in social data. They also present a particle swarm optimisation algorithm to discover
the leading
nodes~(i.e., influential people such as extremist and radicalised individuals) in a social network.
%
To resolve the issue
of recognising the context and ultimate influence of the extremist or radicalised people in social media,
Beheshti et~al.
proposed a context analysis algorithm in their study. They address the issue of detecting those nodes having
a remarkable
effect on boosting the influence of extremist ideas in  social media~\cite{beheshti2021towards}.

\subsubsection{Suicidal Behaviour}

There are several studies leveraging ML and deep learning methods aiming at suicide identification~\cite{coppersmith2018natural, aladaug2018detecting, katchapakirin2018facebook}. Suicide is a global health dilemma, while preventable with timely actions\footnote{https://www.who.int/news-room/fact-sheets/detail/suicide}. It 
is among the most common causes of death
worldwide~\cite{moore2021suicide},
and all the countries experience this issue more or less. The suicide rate in Australian men has grown from 16.2 per 100,000 individuals in 2011 to 18.6 in 2020, while women suicidal rate increased from 5.1 to 5.8 per 100,000 in the same years~\cite{ausgovsuicide}.

Suicide is a cognition-related~\cite{o2008personality} disorder and mostly originated from depression~(i.e.,
two-third
of suicides occurred due to depression). It is also related to some other factors such as the family history of mental disorder, comorbid anxiety and gender~\cite{zou2022core}. Several cognition-related factors could be considered as predictive risk factors of suicide, namely problem-solving impairments, memory and thinking deficiency, negative cognitive style,
the personality trait
of neuroticism, rumination and self-criticism\cite{o2008personality}, causing more discrepancy words usage in their communications~\cite{troop2013expressive}.

In a recent study by Shini Renjith
et~al.~\cite{renjith2021ensemble},
a combined deep learning classification model is leveraged,
aiming at the identification
of suicidal ideation through Reddit ~(i.e., an internet-based social platform) posts analysis. They used a hybrid model,
consisting of
LSTM and CNN for this text classification task. In order to
increase
the model accuracy, they used an attention technique
to mimic
cognitive attention~\cite{jalayer2020attention}.

The reason for using
attention mechanisms
is that some parts of the textual data, which are conveying essential cognitive meaning
could be more relevant and should be utilised.
For example, negative emotion words, more relevant to suicidal behaviour, should be considered more important in data analysis.
Hence, using
an attention-based model
enables researchers to fix different weights according to the words cognitive importance. This weighting, leads to extracting a better emotional sense from
users' posts,
while having more cognitive-related patterns recognised by convolutional neural networks. Their proposed model had 90.3\% accuracy and 92.6\% F1-score, which is greater than the related baseline models~\cite{renjith2021ensemble}.

\subsubsection{Depression and Anxiety Disorder}~\label{Depression-cognitive_impairments}

Depression and anxiety are
two prevalent
mental disorders~\cite{betterhealthstatistics}, being proved to be associated with cognitive impairment~\cite{castaneda2008review}. They are known to be risk factors for some somatic disorders such as cardiovascular disorder (CVD). They're linked to lower quality of life, as well as higher healthcare costs~\cite{angermann2018depression}. They are frequent mental health issues that have significant financial and social consequences~\cite{shafiee2018saffron}.

\begin{itemize}
    \item\textbf{Anxiety}
    As an example instance of mental disorder categories, Anxiety disorders are distinguished
    by a special way
    of thinking and behaviours, associated with ongoing and mostly illogical panic, concern and horror. For example, panic disorder~(i.e., an instance of anxiety disorders)
    is accompanied by
    fear or anxiety attacks that are normally brief but can be so intense that the person believes she/he will collapse or die.
    People who have panic disorder are mainly worried about recurring attacks, and they avoid settings where they may repeat.

    Individuals suffering from social anxiety disorder~(i.e., another instance of anxiety disorders) are mostly afraid of situations where the person believes he/she will be in the middle of attention. They are also concerned about saying shameful things, while at the same time worrying that others will notice their anxiety and criticise them. On the other hand, months of excessive worry about
    day-to-day stuff, eluding or looking for reassurance in
    places
    when the
    result is in question
    are the main thinking pattern of people with generalised anxiety disorder~(GAD). They usually are too anxious about things that could go wrong~\cite{andrews2018royal, edition2013diagnostic}.

    Several studies found that anxious people have decreased performance on a wide range of cognitive tasks.
    Anxious individuals exhibit difficulties with inductive reasoning, decision-making process, memory length, attention control, and inhibition.
    They are
    mainly engaged with emotionally negative concerns. Anxiety may cause people
    to have problems
    with some certain cognitive tasks because they selectively process something irrelevant to what really matters~\cite{dalgleish2000handbook}.

    \item\textbf{Depression}
    Depression~(i.e., a kind of
    depressive disorder)
    is considered to be one of the main
    origins of disability
    globally~\cite{de2013predicting, lemoult2019depression}, and expected to be the first by 2030~\cite{pedrelli2014reliability}. This disorder is a leading cause of disability, with 5\% of adults suffering from it\footnote{https://www.who.int/news-room/fact-sheets/detail/depression}.

    There are several studies demonstrated that
    cognition may have
    a significant role in the start and
    duration of depression~\cite{lemoult2019depression, hammar2009cognitive}. Depression is noticeably associated with cognitive functioning impairments such as adversity in executive functioning, working memory and processing speed~\cite{ahern2017cognitive, chakrabarty2016cognitive}. Depression-related cognitive functioning includes `cognitive control problems', `negative cognitive biases' and `maladaptive cognitive emotion regulation strategies'~\cite{lemoult2019depression}.

    Cognitive control problems, being prevalent in depressed people, means
    the inability
    to prevent negative information from
    going into
    the working memory~(i.e., the current memory). Besides, negative cognitive biases refer
    four types
    of processing as follows. Negative `self-referential' processing, which is indicative of
    having a negative
    self-schema and negative self-descriptions. Negative `attentional' biases, showing difficulty in not paying attention to the negative drivers. Negative `interpretation' which means having tendency to interpret events and information in a negative way. Negative `memory bias', also refers to remembering more negative memories rather
    than positives.

    In addition, there
    are three
    maladaptive cognitive emotion regulation strategies. First, the `reappraisal' strategy, referring to reviewing a past event with the aim of rectifying the related negative feelings.
    A depressed individual
    rarely benefits from this strategy. Second, `rumination', being a maladaptive strategy and one of the depression symptoms. It means thinking about negative past experiences repetitively. And the third strategy is `distraction', being an adaptive strategy, which means causing a person not to think about something.
    Depressed
    people show low levels of distraction from thinking to negative memories~\cite{lemoult2019depression}. Among all mental disorders, in this study, we mostly focus on depression.
\end{itemize}

\subsection{Depression: A Cognitive-Psychological Mental Disorder}~\label{cognitive_science}

Due to the significant
adverse effects
it may have
over global communities, depression has attracted the attention of many researchers and governments to conduct extensive studies in this field. Depression as a cognitive-related disorder is described previously in section~\ref{Depression-cognitive_impairments}. In this section, based on the best practices in the field and the golden standards in cognitive and psychological studies such as 'Diagnostic and Statistical Manual for Mental Disorder'~(DSM) and several other resources~\cite{edition2013diagnostic,lux2010deconstructing,pedrelli2014reliability, de2013predicting}, depression is explored in terms of its symptoms, risk factors, supportive symptoms and the relationships among them.

\subsubsection{Depression Symptoms}

Depression is a complex mental disorder. It contains a set of heterogeneous symptoms~\cite{beevers2019association} such as emotional, cognitive, and behavioural symptoms, with persistent depressed mood and lack of interest in usually enjoyable chores~\cite{kircanski2012cognitive}. In this part, diagnostic symptoms of depression that are mainly used by psychologists
are explained.

\begin{itemize}
    \item\textbf{Depressed Mood:} `Depressed mood' is one of the main symptoms in identifying depressed individuals~\cite{edition2013diagnostic}. As a symptom of
    various
    mood disorders such as depression~\cite{beiter2015prevalence},
    it is determined by some feelings such as
    feeling sad, empty, hopelessness, and discouraged \footnote{https://www.nimh.nih.gov/health/publications/depression/}. Having pessimistic ruminations, being tearful and unhappy could also be indicative of depressed mood~\cite{jang2004heritability, serani2011living}.

    The presence of `depressed mood'
    can sometimes be inferred by facial expression or conduct. Also, depressed mood could be shown by bodily pains and aches,
    a sensation of extreme frustration about little issues,
    and a tendency to react with angry outbursts~\cite{sayar2000anger}. In children and adolescents, this symptom mainly
    may manifest
    itself in the form of irritability rather than feeling dejected or sad~\cite{perlis2009irritability, edition2013diagnostic}.
    \item\textbf{Loss of Interest: }
    Significantly decreased pleasure or interest, at least to some degree, in nearly all of daily activities, and also
    the inability
    to anticipate happiness, almost everyday, is one of the most prevalent symptoms of depression. Individuals may
    feel not caring anymore,
    less interested in hobbies, feel no enjoyment in activities that
    were considered to be
    joyful in past~\cite{endo2022depressive, gilbert2007psychotherapy}. In addition, Social withdrawal\cite{de2013predicting, edition2013diagnostic}, neglect of pleasurable avocations and reduced levels of sexual desire are
    some other
    probable indicators of loss of interest in depressed people~\cite{edition2013diagnostic}.
    \item\textbf{Appetite or Weight Change: }
    Around one-third of depressed people identified with
    an increase
    in appetite, while about
    half
    of them report appetite decrease~\cite{cosgrove2020appetite}. Depression is
    accompanied by
    several physical health problems such as cardiovascular disease (CVD) and obesity~\cite{chapman2005peer}.
    These diseases
    are mainly associated with depression somatic symptoms such as problems with sleeping and fatigue,
    so there
    is a mutually-reinforcing relationship between them. Change in appetite, being a depressive somatic symptom, is also affected by these kinds of relationships.

    A potential biological cause
    that relates
    depression to those physical health issues is systemic inflammation~\cite{milaneschi2019depression, miller2016role, bhattacharya2016role}. Inflammation is defined as the immune system’s reaction to damage or irritation. Stress, infection, or chronic diseases could lead the immune system to create inflammatory responses
    throughout the
    body~\cite{gaucherinflammation}.

    Based on some studies, the different underlying neural and inflammatory factors could cause some depressed people to tend to infer foods to be pleasant
    (i.e., increased appetite). In contrast,
    others feel reluctant to eat most of the foods~(i.e., decreased appetite)~\cite{cosgrove2020appetite}.

    %
    \item\textbf{Sleep Disturbance: }
    Unusual changes in sleep patterns nearly every day
    are common somatic symptoms of
    depression~\cite{edition2013diagnostic}. Hypersomnia is characterised by recurrent episodes of Excessive Daytime Sleepiness~(EDS).
    It's often considered to be the result of interrupted or poor sleep, and it's linked to a variety of sleep problems,
    such as insomnia~\cite{ncbihypersomnia}.

    Insomnia is defined to be
    a sleep-related difficulty
    such as falling asleep or
    remaining
    asleep for a considerable period of time~\cite{davis2014fighting}, waking up during the night and then
    having difficulty
    falling asleep again, and also experiencing increased daily sleep\footnote{https://www.ampsych.com.au/psychology-treatment-sydney/insomnia/}. Insomnia could have reflections such as experiencing irritability, aggression and anger, lack of energy, concentration and motivation, loss of interest in daily activities. It also could lead to some physical tensions such as headaches\footnote{https://www.healthdirect.gov.au/insomnia}.
    \item\textbf{Psychomotor Problems: }
    Two different demonstrations of psychomotor problems are psychomotor agitation and psychomotor retardation. Psychomotor agitation
    is defined as
    engaging in purposeless movements, e.g., pacing around the room, rapid talking, and tapping the toes and feeling restlessness~\cite{edition2013diagnostic}. Also irritability and suicidal attempts are also considered as its symptoms. The United States (US) Food and Drug Administration
    has noted that the inner tension and mental distress are, respectively, the original cause of excessive motor activities and restlessness\footnote{https://www.ncbi.nlm.nih.gov}.

    On the other hand, Psychomotor retardation could be recognised through some symptoms such as Speech slowness, decreased mobility and cognitive dysfunction~\cite{buyukdura2011psychomotor}. Cognitive functions are indicative of several significant mental processes that directly affect our everyday personal and social behaviours. Perceptual processes, memory, decision-making, language comprehension and attention processes are part of
    that cognitive
    functioning~\cite{nouchi2014improving}. Besides, previous studies demonstrated that individuals with negative processing biases~\cite{zhang2022neural, gollan2008identifying, roiser2012cognitive, rude2003negative} and negative cognitive styles are more susceptible to be depressed~\cite{brown1986social}. The negativity bias refers to the situations where negative events such as unpleasant thoughts, emotions, or social interactions influence
    an individual's
    psychological state and processes much greater than equally emotional events but positive~\cite{kanouse1987negativity,rozin2001negativity}.

    \item\textbf{Fatigue: }
    Sustained loss of
    energy and tiredness without
    physical exertion are some of the fatigue symptoms.
    Even the smallest tasks seem
    to be hard to be done and efficiency in doing tasks may be decreased. This feeling is indicated in a large number of depressed cases~\cite{edition2013diagnostic}.
    Fatigue could also increase
    the likelihood of making mistakes.

    Based on medical studies, Insomnia and fatigue are interdependently related to each other while having different essences~\cite{davis2014fighting}. Fatigue is not the mere feeling of tiredness or drowsiness and could be caused by lack of sleep, long physical or mental activities. Making errors,
    decreased productivity
    reduced alertness are some of the probable adverse effects of being fatigued~\cite{ausgovfatigue}.
    \item\textbf{Negative Self-Evaluation: }
    It is very
    usual
    for depressed
    individuals
    to have
    a negative self-image, feelings and thoughts of self-loathing and being valueless during their depressive episodes~\cite{serani2011living, endo2022depressive}.
    The presence of
    delusional or near-delusional guilt is less common but is indicative of greater intensity. Depressed people may have unrealistic negative evaluations of their worth and will ruminate
    on their minor
    past failings. They mainly misinterpret the trivial events as evidence of personal fault and blame themselves for failing in meeting various occupational and interpersonal responsibilities~\cite{edition2013diagnostic}.

    Depressed people often tend to react with self-criticism when facing failure, stress or obstacles in life. Individuals
    with an increased
    level of self-criticism use more discrepancy words in their communications. This could be a reflection of different levels of confidence and certainty in achieving life goals or probably the degree to which one person is able to concentrate on specific aspects of favourable targets rather than unclear or general ones~\cite{troop2013expressive}. Having the negative self-evaluation is closely related to depression and is a powerful mediator between stressful
    life events
    and depression~\cite{zou2022core}.
    \item\textbf{Impaired Ability to Think: }
     Indecisiveness, diminished ability to think or concentrate, being easily distracted and having memory difficulties, especially in older people, are some of the complaints that depressed individuals may have~\cite{edition2013diagnostic}. Indecisiveness  means having difficulty with decision-making in everyday life situations. Relationships, health, and
     job-related issues
     are some examples of those situations. Indecisiveness is considered to be associated with distress in different daily activities~\cite{hallenbeck2021understanding}. On the other hand, depressed people mainly demonstrate a high level of anxiety which in turn makes it hard to make decisions~\cite{indecisiveness}.

    Lack of motivation is one of the reasons that cause depressed people
    to be indecisive.
    Having decision-making difficulties is considered to be a somatic issue. A previous study has demonstrated that depressed people suffer from grey matter loss in their brain. Grey matter is some regions
    in the human brain
    that are involved
    with motivation and decision-making ability, emotions, self-control, memory\footnote{https://en.wikipedia.org/wiki/Grey\_matter}.
    \item\textbf{Suicidal Ideation or Thoughts of Death: }
    Suicide is one of the main public health dilemmas worldwide. It was the fourth important cause of death in young people on 2019, globally. There is about 700,000 suicide commitment annually\footnote{https://www.who.int/news-room/fact-sheets/detail/suicide}. Majority of
    depressed individuals
    are shown to
    have suicidal ideation~\cite{zou2022core, ribeiro2018depression, liu2019stressful}. Not only the fear of dying but also recurrent thinking
    about it and suicidal attempts or plans are all kinds of this symptom that can be presented by a depressed person.

    Depressed individuals concentrate on
    committing suicide because they believe they are unworthy of life
    or since they feel to be unable
    to deal with the depressive anguish~\cite{edition2013diagnostic}. Based on a recent study~\cite{glenn2020can}, people who are approaching suicidal attempts tend to use more anger-related terms while the positive emotion in their communications is reduced. Besides, negative self-evaluation, negative cognitive style are known to be probable risk
    factors for
    suicidal deaths~\cite{o2008personality}

\end{itemize}
An individual would be diagnosed to be suffering from depression if she/he exhibits either depressed mood or loss of interest and
four (or more) of the other
above-mentioned symptoms.
In addition,
this depressive episode must be followed by distress~\cite{stein1995mixed} or a reduction in social or professional performance.
These symptoms must have been present during
at least two
consecutive weeks,
indicating a change in functioning compared to the past.
It also must be noted that those symptoms that are obvious attribute of another medical conditions are excluded~\cite{pedrelli2014reliability, edition2013diagnostic}.

\subsubsection{Risk Factors and Supportive Symptoms}

\textbf{Risk Factors.}

Depression risk
factors include the type of
personality traits that a person have, demographic features, having previous episodes of depression, and having major non-mood disorders.
In this part, first,
personality traits and the way they could affect potential people to be depressed are described. Then other risk factors such as demographic features would be explained.
\begin{itemize}
    \item\textbf{Personality Traits: }\label{P_Traits}

    Suffering from depression
    is linked with some of the personality traits~\cite{cloninger2006can}. Personality is recognised by the way people tend to behave, react to different daily events, and also their patterns of cognitive and emotional processes~\cite{jaiswal2019automatic,yakhchi2020enabling}.
    A wide range of past and recent research is evidence of personality traits' effect on vulnerability to some mental disorders.
    It is also known that personality is a significant risk factor of depression~\cite{kendler2007sources,kendler2006personality,klein2011personality,ramirez2021relation,lyon2021associations,li2020linking,alizadeh2018predictive,allen2018big,hakulinen2015personality,bagby1995major}, and has been hypopapered to be a depression predictor~\cite{hakulinen2015personality}.

    Based on The most acceptable and universal personality model,
    the Five Factor Model (FFM),
    there are
    five different
    personality traits, named ``Big Five”. These Big Five are `Neuroticism', `openness', `Extraversion', `Agreeableness', and `Conscientiousness'~\cite{klein2011personality, allen2018big}.
    Extraversion is defined to be
    genial, chatty, assertive, and energetic, while neuroticism means to have negative affectivity, be uneasy, sad, irate, and unsure. On the other hand, conscientiousness trait the characteristics such as being careful, perfect, accountable, and  orderly. Agreeableness also is indicative of traits such as being kind-hearted, compliant, unselfish,
    and gentle.
    Lastly, openness is defined as
    being creative and open-minded~\cite{roccas2002big}. The main personality traits that are considered to be depression risk factors are neuroticism, extroversion, and conscientiousness~\cite{hakulinen2015personality, allen2018big}.

    Neuroticism as the most significant trait~\cite{kendler2007sources, kendler2004interrelationship, kendler2006personality} is positively correlated with depressive symptoms. The other two, i.e., extraversion, and conscientiousness,
    have a negative
    correlation with susceptibility to be depressed~\cite{klein2011personality}. Neuroticism, contains two sub-domains, namely `withdrawal' and `volatility'. Extraversion consists of `enthusiasm' and `assertiveness', and conscientiousness also includes `industriousness' and `orderliness'. Sadness and anxiety are two examples of Withdrawal referring to internal negative emotions, while volatility
    refers to
    external negative emotions such as irritability and anger. On the other hand, enthusiasm and industriousness are negatively correlated with depression~\cite{allen2018big,lyon2021associations}. Hence, neuroticism could be viewed as a risk factor while extraversion and conscientiousness are considered to be protective factors~\cite{lyon2021associations, alizadeh2018predictive}.

     Neuroticism is considered to be
     the state of having negative affectivity~\cite{edition2013diagnostic}, and
     it has been
     shown that depressed individual have higher negative emotions than not-depressed individuals\cite{de2013predicting}. The word “affect” refers to
     our experienced emotions or feelings
     and the way they influence us in making decisions~\cite{posandnegaffect}. Negative affectivity
     relates to a variety
     of negative emotions and poor self-concept, that consists of symptoms such as distress, sadness, disgust, lethargy, and fear~\cite{koch2013can}. Distress is a kind of negative stress in comparison with
     eustress, which is positive stress.
     Distress has characteristics such as causing anxiety, experiencing unpleasant feelings, reducing the performance, resulting in physical or mental problems~\cite{EustressDistress}.
     Lethargy is a condition in which
     an individual mostly has
     feelings of being fatigue, mood change, reduced energy, alertness or ability to think\footnote{https://www.healthline.com/health/lethargy}.

    \item\textbf{Demographic Feature and Other Risk Factors: }

    Based on previous studies, the prevalence of depression differs in different age groups.
    It is three times higher in 18- to 29-year-olds than in 60-year-olds~\cite{edition2013diagnostic}.
    Females in their early teens had 1.5 to 3-fold higher levels of depression~\cite{kendler2004interrelationship, essau2010gender, cosgrove2020appetite}. It is also shown that
    there are no obvious variations in symptoms, duration, therapeutic response, or outcomes between men and women.
    Besides
    demographic features such as age and gender, there are other risk factors being effective in suffering from depression.

    Having a previous history of depression also could make individual more susceptible to be depressed again in future~\cite{fried2015depression, kendler2004interrelationship}. In addition,  having the background of other non-mood disorders such as substance-related disorders, prominent anxiety~\cite{jang2004heritability, hettema2006population}, borderline personality disorders as the most common ones, as well as
    Other depression risk factors include disabling medical illnesses such as diabetes, severe obesity, and heart disease~\cite{edition2013diagnostic}.
\end{itemize}
\textbf{Supportive Symptoms:}
\begin{itemize}

    \item\textbf{Absolute Word usage : } Depressed individuals mostly talk in
    the absolute language and use a greater
    percentage of absolutist words in their natural language and daily communications~\cite{serani2011living,al2018absolute}. Absolutist thinking is the cause of several cognitive falsifications and illogical thoughts that are said to be the mediators of core emotional disorders.
    There are 19 absolutist words that have been properly verified,
    e.g.,
    `definitely', `absolutely' and `never'~\cite{al2018absolute}.
    %
    \item\textbf{Abnormal Fear : } As mentioned in section~\ref{P_Traits},
    fear is one of the behavioural symptoms of neurotic individuals.
    It is also illustrated in
    American Psychiatric Association Manual~\cite{edition2013diagnostic} that having abnormal fear and phobia could be a supportive symptoms of being depressed.
    \item\textbf{Tearfulness : } Crying is a powerful sign of organismic distress
    and acts as distress reduction.
    Clinical studies have indicated that depressed people are crying excessively~\cite{beck1979cognitive} which may reflect the intensity of distress~\cite{rottenberg2003vagal}. American psychiatric association also has introduced being tearful as a supportive symptom for identifying depressed individuals~\cite{edition2013diagnostic}.
    \item\textbf{Physical Pains : } Being depressed and having physical pain are strongly connected~\cite{edition2013diagnostic} and could make a vicious cycle.
    Back pain or headache is
    the first and most common unexplained
    body pain, caused by depression.
    Pain and its consequent issues could affect individuals' mood. Chronic pains could lead to depression, whether injury-related ailments or pains originated from a physical disorder such as diabetes, cancer or heart disease~\cite{depandpain}.
    \item\textbf{Obsessive Ruminations/Concerns : } Obsessive ruminations or thoughts are mainly demonstrated in depressed people~\cite{edition2013diagnostic}.
    Rumination is defined as
    thinking frequently over a single or specific thoughts which are inducing negative feelings. Obsessive rumination could be harmful in terms of mental health. It causes difficulties
    in the ability to
    think and emotion processing. Also,
    Choudhury et~al. illustrated
    in their study that depressed people tend to use more relational and medicinal concerns\cite{de2013predicting}.

    Having obsessive ruminations, also could make the depression more intense or prolonged. It may cause feeling alone and
    isolated, which in turn causes
    pushing people away\footnote{https://www.healthline.com/health/how-to-stop-ruminating}. There are some reasons for rumination, e.g., experiencing previous physical or emotional trauma, dealing with continuous uncontrollable
    stressors. Also,
    Neurotic people often tend to ruminate over their relationships~\cite{aparummination}.
    \item\textbf{Self-focused Attention : } Several previous studies have shown that people with depressive symptoms tend to use special words while talking or writing. It is recognisable from the way individuals express their thought whether they are suffering from depression or not. Depressed people use more first-person pronouns, i.e., `I', `Myself', `Me', showing an increase in self-focused attention~\cite{de2013predicting,serani2011living,tausczik2010psychological,rude2004language}. Based on depression cognitive models, self-referent attention and processing, which is negatively biased, have important roles in suffering from depression~\cite{beevers2019association}.
    These findings are in line
    with the fact that one of the symptoms of depression is negative self-evaluation and feeling of worthlessness~\cite{edition2013diagnostic}. Hence, using self-focused words while they are blaming or criticising themselves seems very natural.
    \item\textbf{Anxiety : } Anxiety is a natural body reaction to stressful events.
    It is accompanied by
    the fear of what’s to come. For example, the first school day or having a job interview could cause anxiety and nervousness.

    Anxiety and depression happen independently, however, they could be considered as comorbid disorders that happen at the same time. Anxiety could be a supportive symptom for depression~\cite{edition2013diagnostic}, which leads other symptoms to be worsened\footnote{https://www.healthline.com/health/anxiety}.
    It is also
    considered to be a probable symptom of negative emotion~\cite{allen2018big,lyon2021associations}. Although it could be a kind of background non-mood disorder~\cite{edition2013diagnostic, jang2004heritability, hettema2006population}, anxiety is mostly illustrated in
    depressed people's
    behaviour~\cite{indecisiveness}.
    \item\textbf{Irritability : }
    Irritability is known to be the state of quickly becoming annoyed, angry and frustrated even with minor issues~\cite{de2021predictors}. Irritability is considered to be one of the supportive symptoms in depression identification. Depressed individuals,
    especially children,
    might feel irritated frequently. Sadness and depressed mood in Children and adolescents mostly is demonstrated as irritability, so that they would react with anger to slight provocation~\cite{edition2013diagnostic}.

    Other symptoms of irritability could be feeling restlessness, difficulty concentrating and agitation. On the other hand, there are some causal mental and physical condition such as stress, sleeping problems, chronic pain and thyroid that makes individuals experience irritability\footnote{https://www.healthdirect.gov.au/irritability-and-feeling-on-edge}.

    \item\textbf{Special Phrases : } Some of the most frequent phrases that can be seen as signals for being depressed are ``I should", ``I can’t…", ``It’s all my fault", ``I’m tired", ``I want to be alone", ``No one cares". As some may believe it is a stigma
    to have a mental disorder,
    depressed people often try to hide it behind their positive phrase that ``I am fine". Two signalling phrases of losing interest in normally enjoyable activities are ``I don’t feel like it" and ``It’s not fun anymore". Besides, ironically use of ``I feel better" and also saying ``What’s the point?” could be warning since all of those mentioned feelings can become overwhelming and make people want to give up~\cite{serani2011living}.

    As mentioned before in section~\ref{P_Traits}, neurotic people are more likely to be depressed, while showing more negative emotion in their daily life. Also, depressed people are more likely to use negative emotion terms than not-depressed people~\cite{thompson2021emotion, tausczik2010psychological}. The following words are some examples of negative emotion terms could be used by some of depressed individuals, `down', `stressed', `upset'
    ~\cite{serani2011living}.

    \item\textbf{Depression Unigrams : } Based on a study by
    Chouldhary et~al.,
    depressed people tend to use special depression unigrams and terms in their communications. As Figure~\ref{fig:depression_unigrams} shows, depression unigrams are categorised into 4 areas. Treatment and
    relationship, life
    words refers to medicinal and relationship concerns, respectively.
    These are concerns assumed
    to be related to obsessive ruminations, a previously mentioned supportive signs. Also, symptom and disclosure words are considered to be
    separate supportive symptoms
    for depression~\cite{de2013predicting}.

\end{itemize}
\begin{figure}[t]
    \centering
    \includegraphics[width=0.5\textwidth]{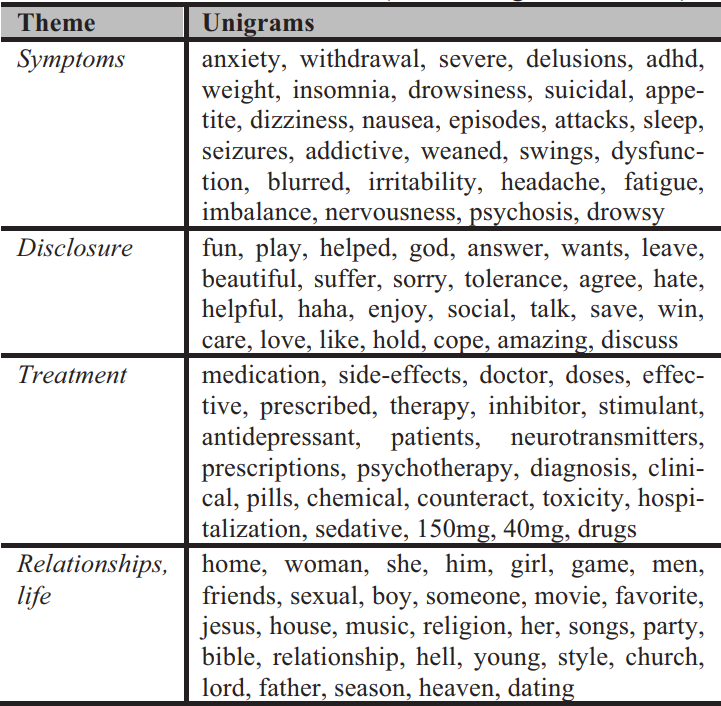}
    \caption{List of depression unigrams that depressed people tend to use in their communications as indicated in Chouldhary et~al. study~\cite{de2013predicting}.}
    \label{fig:depression_unigrams}
\end{figure}

\subsection{Depression Identification/Assessment Approaches}~\label{Identification_Assessment_Approaches}

Likewise, with other
chronic disorders, the sooner treatment of depression will help the full remission of it~\cite{edition2013diagnostic}. Therefore, due to the importance of this issue, a valuable collection of studies has been done over depression from various perspectives.
In this section,
we will review three different approaches~(i.e., clinical, non-clinical and knowledge-based), aiming
at the identification
and assessment of depression.

\subsubsection{Clinical Approaches}
\label{Assessment_Tools}

In contrast to most physical disorders that can be identified through methods such as imaging and laboratory tests,
the majority of
psychological and cognitive abnormalities are not easily diagnosed. Generally, mental health issues need to be assessed before any Therapeutic prescription~\cite{gorenstein2021assessment}. Depression, considered to be
the main
health problem around the world~\cite{park2021useful}, is no exception
and attracts
the attention of many researchers in terms of developing reliable assessment gadgets. In order to provide adequate care, validated scales must be used in the screening, diagnosis, and measurement of therapeutic efficacy in depression. Clinicians who learn how to utilise those scales can increase diagnostic accuracy, save time, offer greater constant treatment, and track a patient's complicated emotional and behavioural responses to therapy~\cite{anderson2002depression}.

More than 40 years ago, symptom-based scales were established
to give a numeric value
to a wide range of patient behaviours,
emotions,
and feelings. They've subsequently evolved into a confusing array of tools developed for a number of objectives, some of which are fairly broad in scope and others more specific~\cite{anderson2002depression}. Depression assessment tools include screening, diagnosis and monitoring instruments, each designed for a specific aim. To identify potential depressed individuals, short self-reported tools, namely screening measures are used.
On the other hand,
diagnostic tools are interview-based measures used by clinicians. Also, monitoring tools are rating scales to appraise the intensity of depression and identify any changes in the depressive symptoms~\cite{gorenstein2021assessment}.
\begin{itemize}

    \item\textbf{Screening Tools~(BDI-II, CES-D and PHQ-9) :}

    The Beck Depression Inventory~(BDI) is among the most
    commonly
    used screening tools for identifying depression severity~\cite{beck1996beck, gorenstein2021assessment,smarr2011measures,beck1961inventory,whisman2000factor,biracyaza2021psychometric}. The assessment includes questions related to depressive symptoms related to depressive symptoms like sadness, as well as impaired cognition such as feeling worthlessness, and also bodily symptoms like weight loss~\cite{mcdowell2006measuring}. It was first developed in 1961, having 21 items to assess depression-related symptoms. Individuals must answer to multiple-choice questions based on their feelings on the same day as they are answering~\cite{beck1961inventory}.

    In 1971 BDI-IA, being the amended version
    of the original BDI,
    was published. BDI-IA must be responded based on the
    preceding week's events
    and feelings, including today~\cite{beckdepression}. The BDI-II is a modified edition of the BDI that's been released in 1996 to comply with the new version of DSM~\cite{bell1994dsm} depression criteria. It replaced four of the original BDI's items, 
    i.e., `Weight Loss', `Body Image Change', `Somatic Preoccupation' and `Work Difficulty'
    with four new ones
    , namely `Agitation', `Worthlessness', `Concentration Difficulty' and `Loss of Energy'.
    BDI-II must be answered based on the preceding two weeks~\cite{beck1996beck}. Figure~\ref{fig:BDI-II} demonstrates part of the BDI-II\footnote{https://naviauxlab.ucsd.edu/wp-content/uploads/2020/09/BDI21.pdf}.
     \begin{figure}[t]
        \centering
        \includegraphics[width=1.0\textwidth]{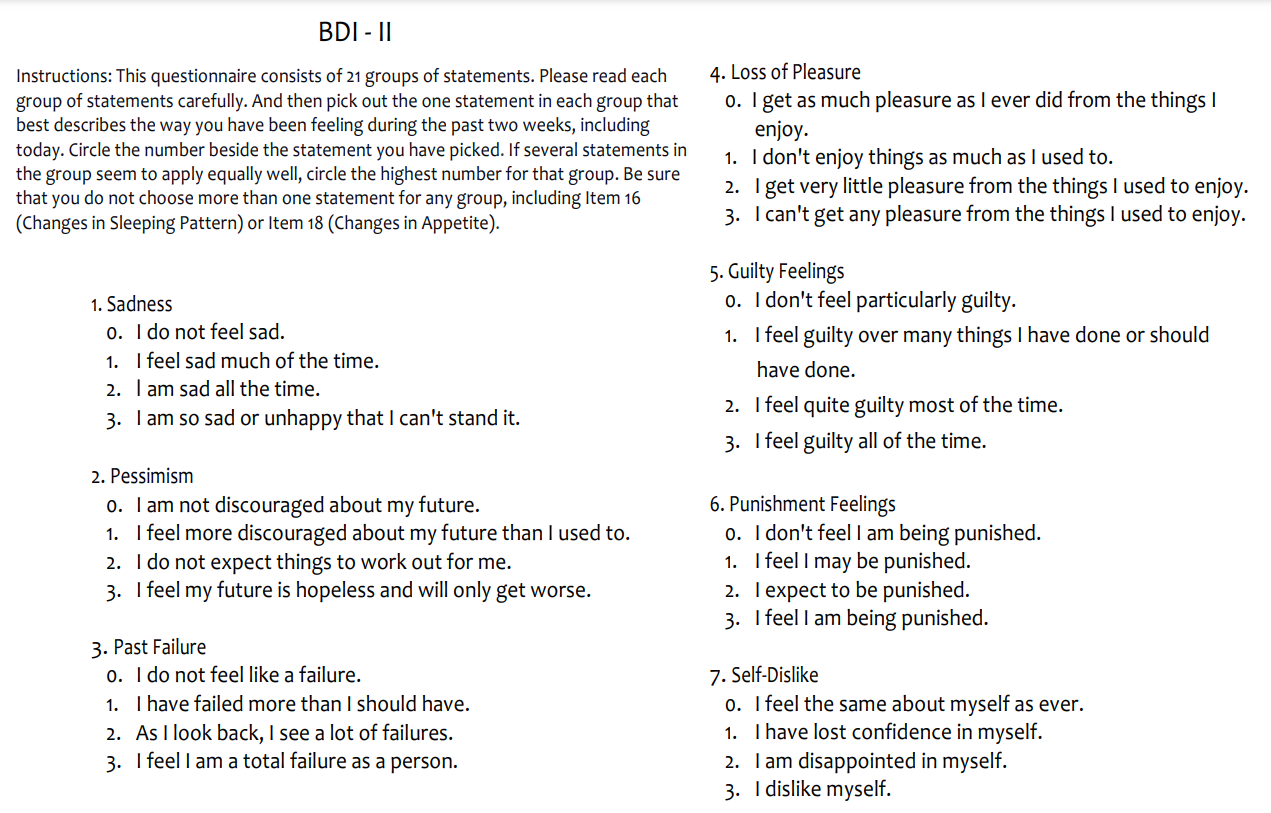}
        \caption{Part of the BDI-II: An extensively used screening tool for identifying depression severity. The assessment includes questions related to depressive symptoms like sadness, as well as impaired cognition such as feeling worthlessness~\cite{beck1996beck}.}
        \label{fig:BDI-II}
    \end{figure}
    The Center for Epidemiological Studies-Depression~(CES-D)~\cite{radloff1977ces, park2021useful, smarr2011measures, carleton2013center,lopez2021self} and the Patient Health Questionnaire (PHQ-9)~\cite{kroenke2002phq, gorenstein2021assessment, dadfar2021patient,costantini2021screening} are 2 of the most important tools for depression screening. As Figure~\ref{fig:CES-D} shows~\cite{CES-D}, CES-D is a 20-question self-report questionnaire.
    It was developed in 1977
    by Laurie Radloff to
    assess depression
    symptoms. This questionnaire should be answered based on the feeling and affections referring to a week prior to the completion day~\cite{radloff1977ces}. Respondents rate each question based on the happening frequency. The rating scale is originated from
    the Likert scale,
    varying between
    0~(i.e., hardly ever or at all) to 3 (i.e., generally or always)~\cite{carleton2013center}.

    Figure~\ref{fig:PHQ-9} also illustrates the PHQ-9 questionnaire, being the most trustworthy in depression screening tool leading to the more accurate identification~\cite{costantini2021screening,abu2015preliminary,ghafari2018social}. The PHQ-9 is frequently used in physician's and therapist's workplaces as part of routine screenings or to assess the mental well-being.
    This simplified version is inspired by the original PHQ form, which includes a wide range of mental health problems such as depression, panic disorder, anxiety, and difficulty sleeping,
    and more~\cite{phq-9}. It is a 9-question tool, each question refers to one of the criteria indicated in DSM-5~\cite{edition2013diagnostic} manual and scores them as `0', `1', `2', or `3'. In which `0' means not at all and `3' means nearly every day \footnote{https://www.ncbi.nlm.nih.gov/pmc/articles/PMC1495268/}.
    \begin{figure}[t]
        \centering
        \includegraphics[width=0.7\textwidth]{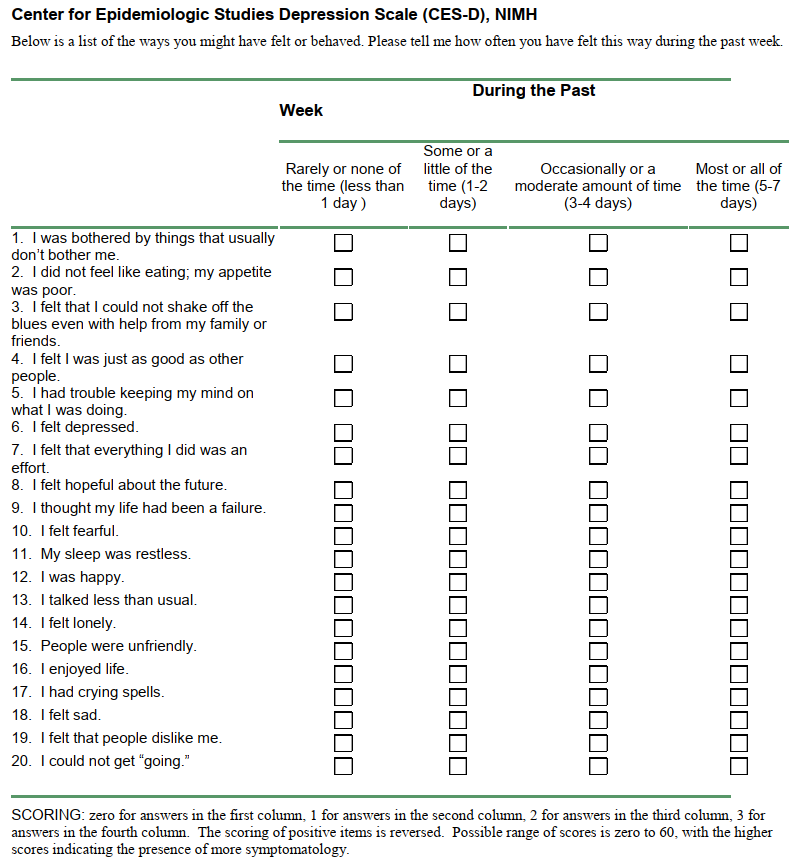}
        \caption{The Center for Epidemiological Studies-Depression: a reliable depression-related questionnaire. It is applied in scientific studies and clinical settings as a depression screening tool.
        It was developed in 1977
        by Laurie Radloff~\cite{radloff1977ces}.}
        \label{fig:CES-D}
    \end{figure}
    \begin{figure}[t]
        \centering
        \includegraphics[width=0.7\textwidth]{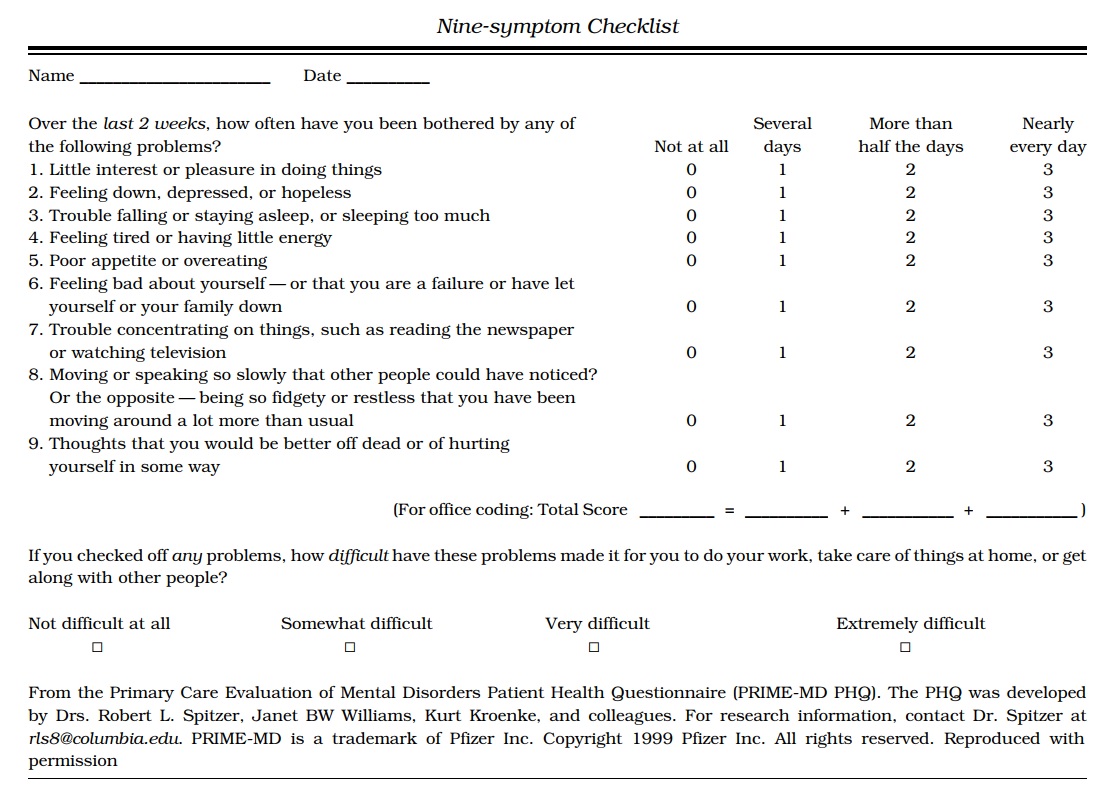}
        \caption{PHQ-9: A 9-question depression screening tool leading to the more accurate depression identification. It is originally developed and published  by Kroenke et~al., on 1999~\cite{kroenke2002phq, kroenke2001phq}.}
        \label{fig:PHQ-9}
    \end{figure}
    \item\textbf{Diagnostic Clinical Interviews~(SCID and CIDI) :}

    Screening attempts are mainly followed by diagnostic interviews. Interviews are used by clinicians to identify potential
    people with
    mental disorders~(e.g., depression). Two reliable mental disorder diagnostic interviews are the `Structured Clinical Interview for DSM Disorders'~(SCID) and The `Composite International Diagnostic Interview'~(CIDI)~\cite{gorenstein2021assessment}. The last version of SCID interview is SCID-5, constructed in accordance with DSM-5 diagnostic criteria. To help with evaluating each criterion as existing or absent, interview questions are simply supplied alongside each relevant DSM-5 criterion~\cite{SCID-5}.

    In addition, the CIDI is a detailed and thoroughly structured diagnostic interview for assessing mental disorders. CIDI, the first and the second versions~(i.e., v1 and V2) were developed by
    WHO in
    1994~\cite{WITTCHEN199457} and 1998 respectively~\cite{andrews1998psychometric}. The latest version is structured in 14 sections, one of them
    especially
    related to depressive disorders. Figure~\ref{fig:CIDI_depression} is the illustration of a segment in CIDI-V2, being associated with depressive disorders identification interview~\cite{CIDI-V2}.
    \begin{figure}[t]
        \centering
        \includegraphics[width=0.7\textwidth]{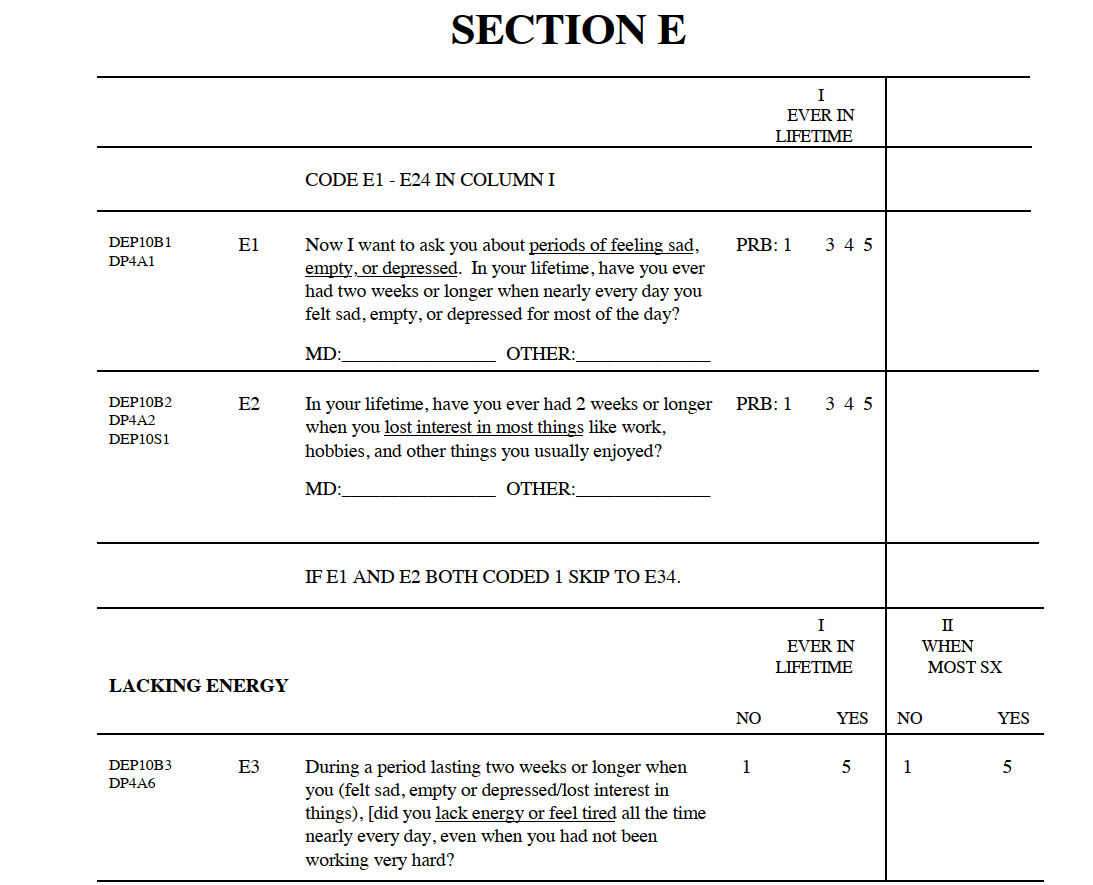}
        \caption{A segment of Composite International Diagnostic Interview~(CIDI) related to depression identification interviews. CIDI is a detailed and fully structured diagnostic interview for assessing mental disorders~\cite{andrews1998psychometric}.}
        \label{fig:CIDI_depression}
    \end{figure}
    \item\textbf{Monitoring Tools~(HAM-D, MADRS) :}

    Monitoring tools are rating scales to appraise the intensity of depression and identify any changes in the depressive symptoms~\cite{gorenstein2021assessment}. Two valuable supplementary tools to be used following a proper diagnosis are the `Hamilton Rating Scale for Depression'~(HAM-D)~\cite{galvao2021potential} and the `Montgomery-Asberg Depression Rating Scale'~(MADRS)~\cite{gorenstein2021assessment}. HAM-D is
    commonly used measure
    for assessing the level of depression in depressed individuals.

    The first unstructured version, HAM-D21 was published by Max Hamilton in 1960 and composed of 21 components.
    Over the years, different versions of the HAM-D have been developed and used~\cite{Hamilton56, carrozzino2020hamilton}. Being structured~(i.e., containing complementary interview questions) or not is the main difference among those versions.
    Due to some limitations, such as inefficiency in detecting changes in the course of treatment, criticisms have been levelled at unstructured types.
    Hence, MADRS, being a structured scale, was
    developed in 1979
    by Montgomery and Asberg to improve the sensitivity of HAM-D~\cite{montgomery1979new}.

\end{itemize}

\subsubsection{Non-Clinical Approaches}~\label{computer_science}

In addition to the clinical approaches explained before, which use assessment tools for screening and identification, computer science society also did a great job in this field by developing various automatic algorithms for identifying or predicting depression. Depression has been investigated through several approaches. Each of those approaches is a special and analysable aspect of depression. For example, facial gestures or language use of depressed people are two aspects that could be analysed. In this way, various data types
have been analysed
in previous researches.
Based on the literature,
these types of data include medical data~(e.g., fMRI  records and EEG), facial, video, vocal, and social media data, In which social media is a rich source of textual data.

There are numerous
modern
ML techniques and classifiers
that have been used
for analysing these data~\cite{kumar2020assessment}. The followings are some of the example techniques leveraged in computer science-based analysis:
NLP, convolutional neural networks~(CNN), ML classifiers such as Support Vector Machines~(SVM), Naive Bayes~(NB), Decision trees~(DT), Multi-layer Perceptron~(MLP), Logistic Regression~(LR), etc. Table \ref{tab:Dep_Idendification_Approaches} is demonstrating different approaches in this regard.
\begin{itemize}
    \item\textbf{Medical Data}\label{medical_Data}
    \begin{itemize}
        \item{FMRI Records: } As we mentioned before, some depressive symptoms could be derived from zonal anomalies in brain blood flow and activities~\cite{wikifmri}. Functional Magnetic Resonance Imaging~(fMRI) measures these activities and changes~\cite{magneticresonance}. Hence, several past studies have used fMRI signatures and leveraged different ML techniques for their studies on depression identification~\cite{fu2008pattern,zeng2014unsupervised}.These capabilities are arising from ML algorithms' power to contract several variables, e.g., MRI signals in brain voxels\footnote{Voxel is a 3-dimensional unit that embeds the signals in brain scans}, into a scale
        that takes
        the intricate patterns derived from these variables~\cite{sato2015machine}.

        Fu et~al.~\cite{fu2008pattern} used SVM technique to analyse fMRI data, and Zeng et~al.~\cite{zeng2014unsupervised} have developed a maximum margin clustering-based unsupervised ML model with 92.5\% group- and individual-level clustering consistency, That model was used for identifying depression through analysing resting-state fMRI scans in the absence of clinical information. Based on their study, some brain functional connectivity networks may have a serious role in suffering from depression.

        Previous research has revealed that
        people who previously suffered from depression are more vulnerable to be depressed again in future~\cite{eaton2008population}. Similarly, the fMRI data have enabled researchers to do different studies on treated depressed patients aiming at identifying imaging bio-markers of depression susceptibility. For example, Sato et~al.~\cite{sato2015machine} have developed a ML algorithm, Maximum Entropy Linear Discriminant Analysis~(MLDA) with 78.26\% accuracy in this regard.

       \item{Electroencephalography Signals: }
       EEG is an electrophysiological measuring technique for recording electrical impulses on the scalp, which has been proven to correspond to the macroscopic activity of the brain's surface layer beneath.
       EEG measures and record voltage changes
       arising from an electrical current within the brain's neurones~\cite{wikiegg}. Several past researches have shown that using ML techniques to analyse EEG recorded data and signals would lead to outstanding results in terms of identifying the predominant mental condition of patients and depression identification.

       Betul Ay et~al.~\cite{ay2019automated} and Sharma et~al.~\cite{sharma2021dephnn}, both leveraged CNN and LSTM models in their studies. The first resulted in classifying depressed cases with the accuracy of 99.12\%  for the right hemisphere signals analysis and 97.66\% for the left hemisphere. The second study has been done with 99.10\% accuracy and the mean absolute error (MAE) of 0.2040. In addition Aydemir et~al.~\cite{aydemir2021automated} used k nearest neighbor~(kNN) classifier with 99.11\% and SVM with 99.05\% accuracy, and Seal et~al.~\cite{seal2021deprnet} apply a CNN model with the accuracy of 0.914
       to classify EGG data into
       two classes, namely related to the depressed or related to not depressed individuals.
    \end{itemize}
    %

\begin{table*}[htbp]
    \centering
    \includegraphics[scale=0.5]{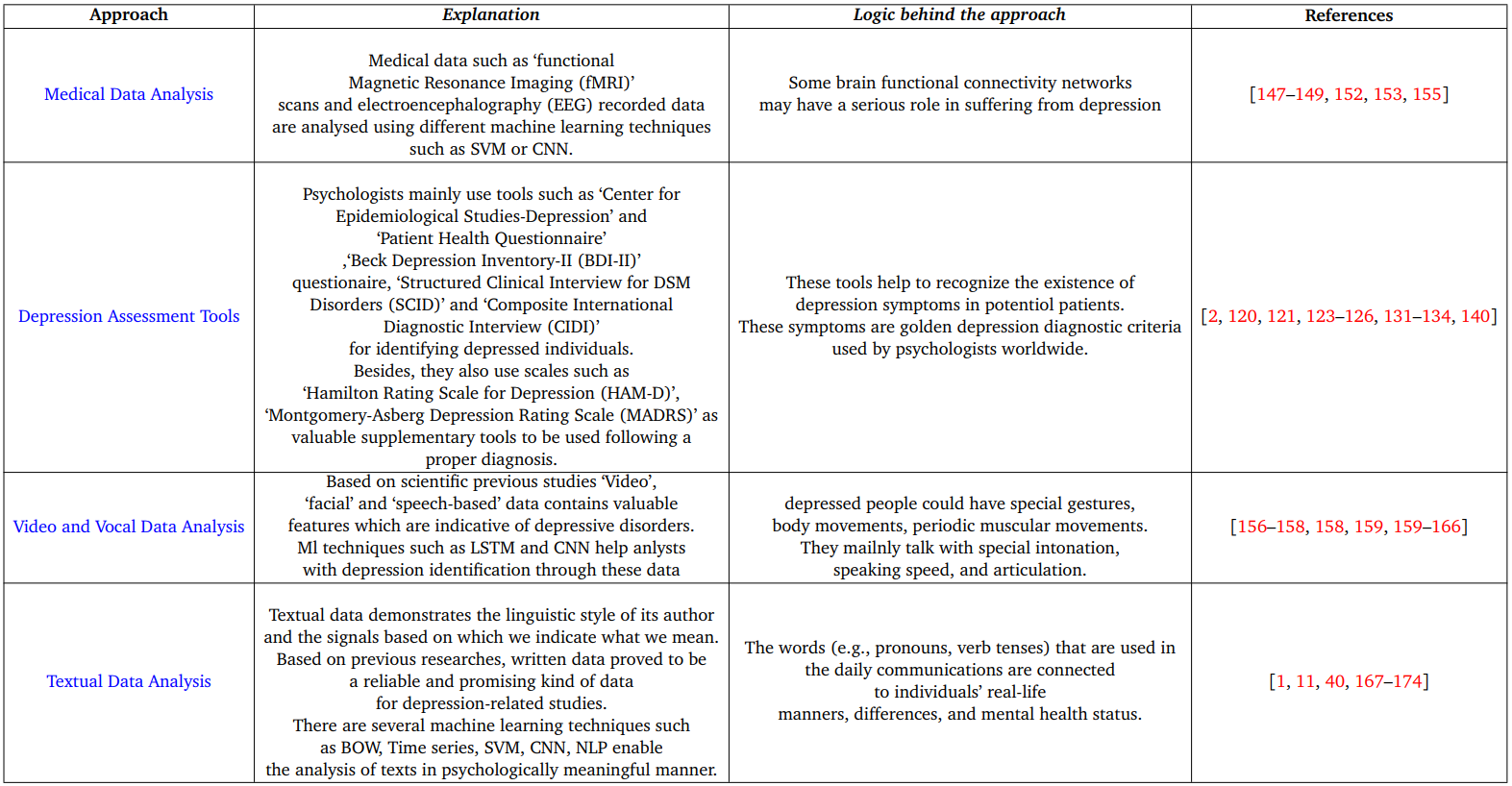}
    \caption{Depression Identification/Assessment Approaches.}
    \label{tab:Dep_Idendification_Approaches}
\end{table*}
    \item\textbf{Video and Vocal Data}\label{Video_vocal_Data}

    There is a growing interest
    in analysing audio and video data to create any probable advances in depression-related areas.
    Since the facial expression could be used vastly as a depressive disorder indicator~\cite{ctx42295508060002171}, there are numerous image- and video-based depression identification studies done previously\cite{narayanrao2020analysis, he2021intelligent}
    to help psychologists with depression diagnosis and therapy~\cite{sun2017random}. Detectable video features such as gestures and
    movements, and periodic
    muscular movements make video data analysis a useful approach for depression studies~\cite{yang2016decision}.

    Image- and video-based depression identification studies leveraged a wide range of automatic modern tools. Deep regression network with focusing
    on a single
    facial image~\cite{ctx42295508060002171}, Deep Learning fed with facial images sequence from videos~\cite{he2021depnet} are some examples of those tools~\cite{niu2020deep}.
    In addition,
    3-dimentional convolutional neural Networks used over video clips having access to spatiotemporal feature aggregation module (STFAM)~\cite{he2021intelligent}, and also CNN models with attention mechanism using video clips as input data~\cite{he2021automatic} are conducted with acceptable performance.

    In addition to video and facial images, there are several vocal- or speech-based studies to developing computerised depression identification models. Generally, depressed people tend to talk slowly and communicate using short sentences in a monotonic way~\cite{williamson2013vocal,cummins2011investigation}. Vocal data could enable identifying those depressive
    patterns by
    analysing the intonation, loudness, speaking speed, and articulation~\cite{sun2017random,yang2016decision}.

    The work of Srimahhur et~al.~\cite{srimadhur2020end}
    is an example of Vocal data-based studies. In this work spectrogram-based and also end to end CNN models are used with the aim of identifing potential depressed people.
    Xingchen Ma et~al., also
    developed DepAudioNet, a deep model made up of CNN and Long ShortTerm Memory (LSTM),
    enabling the classification of
    depressed cases through encoding depressive patterns in the vocal channel~\cite{ma2016depaudionet}.
    Samareh et~al.~\cite{samareh2018detect}, also
    did an integrative
    multi-modal
    study in which, they used the audio, video, and also textual
    data, in addition to
    applying gender variety in their research. Their study had promising results in terms of depression symptom identification.
    \item\textbf{Textual Data}\label{Text_Data}

    These data demonstrate
    the linguistic style of its author and also any other possible metadata. Linguistic style refers to a group of signals based on which we indicate what we mean, and also interpret other people's meaning. It includes features like speaking rate and loudness, being straightforward or not, the word choices~\cite{tannen1995power}, etc.
    The linguistic style could be reflected in
    the spoken and written languages, and the data derived from both could be collected, stored and analysed.

    Based on previous researches, written(i.e., textual) data proved to be a reliable and promising kind of data for depression-related studies, through analysis of different words in sentences and documents~\cite{chafe1987relation}. Kinds of words~(e.g., pronouns, verb tenses) that are used in the daily communications are connected to individuals' real-life manners, differences, and mental health status. In comparison to mentally healthy people, depressed people
    frequently utilise the first person pronouns~
    (i.e., self-focused words), negative
    emotion, and sometimes
    more death-related words~\cite{tausczik2010psychological}.

    One of the main and golden standard computerised text analysis tools is the Linguistic Inquiry and Word Count~(LIWC) program. It is leveraged as one of the most beneficial ways of extracting various word-based features from textual data such as daily diaries, school assignments, and journal articles~\cite{pennebaker2015development, tausczik2010psychological,pennebaker1999linguistic}. In analysing textual data, LIWC counts words in psychologically meaningful classes namely function and emotion words. These classes are indicative of emotional and social conditions, motivations, purposes, and thought patterns. Hence, all the LIWC's capabilities make it a suitable source for personality, behavioural, and mental health related studies~\cite{tausczik2010psychological}.

    The LIWC lexicon was created to analyse a variety of text types, including e-mails, speeches, poetry, or captured daily conversation. LIWC has 90 linguistic, behavioural, and psychological dimensions, consists of 41 word categories (i.e., a list of words) that
    use psychological notions
    (e.g., affect, cognition, biological processes, and drives), and six personal concern categories (e.g., 
    job,
    home, leisure activities), etc~\cite{pennebaker2015development}. These lists were made with the help of 
    emotion assessment methods and glossaries, and a judging panel validated them~\cite{golder2011diurnal}.
    Various studies leveraged
    LIWC as a reliable tool for emotion and linguistic analysis~\cite{de2013major, de2013predicting, golder2011diurnal, saha2022social, boyd2021natural}. It is a valuable resource that is widely utilised for measuring positive and negative affect~\cite{de2013major, golder2011diurnal}, while being a well-known lexicon used in content analysis~\cite{kapoor2007selective}.
    There is also another lexicon
    used aiming at emotion and affect analysis, such as ANEW~(Affective Norms for English Words)~\cite{bradley1999affective, nielsen2011new}, NRC~\cite{mohammad2017word, mohammad2013crowdsourcing}, Warriner Lexicon~\cite{warriner2013norms}, and MPQA~\cite{wiebe2005annotating}.

    Recently, considering the
    popularity of
    daily communications, social media networks
    turned out to be
    a great textual data source for identifying factors associated with depression. This identification, in turn, is a great work in terms of helping psychologists, families and societies to deal with depression side effects. On the other hand, this identification is empowered by different ML techniques'
    capabilities that make
    the textual data analysis possible~\cite{tsugawa2015recognizing, de2013predicting}.

    Social media networks are platforms with multifaceted applications. For example, Twitter, Facebook, and Reddit are common spaces in which people
    can share
    their beliefs, ideas, thoughts, feelings, experiences, and also many
    can find answers to
    their questions. Previous studies show that these networks could help with screening society tendencies and health-related issues such as depression~\cite{chiong2021textual}.
    Several scientific studies have
    analysed social media data using several ML algorithms with the aim of depression identification and prediction, some of them are as follows.

    In a leading study by De Choudhury et~al. (2013), a classifier was built to recognise depressed Twitter users based on their social media activities. They introduced social media as a data source for identifying symptoms of depression in a user. They used different features for their analysis namely, user engagement and emotion, depressive language use, linguistic style, ego-network characteristics such as followers and followees number, and mentions of antidepressant medications.

    Based on this study, depressed people have notable self-focused attention, lower social activity, higher negative emotion, raised medicinal and relational concerns and increased indication of religious thoughts.
    In addition, depressed individuals
    tend to have virtual activities mainly after 8:00~pm or in the early morning, while non-depressed people prefer to be active on online social media, after work-time and mainly early nights. In this study, several classifiers were tested and SVM, with an average accuracy of 70\% and precision of 74\% for depression class resulted in the best outcomes~\cite{de2013predicting}.

    In 2015, Tsugawa et~al. identified a method to show the depression level in Twitter users. They used past activities of users to make different classifiers among which SVM had the best performance with 61\% precision and 66\% accuracy. The main features used in their model were word frequencies, tweets' topic , ratio of positive-affect and negative affect words, tweets' daily frequency, average number of words per tweet and several
    metadata
    such as number of
    followers~\cite{tsugawa2015recognizing}.

    Orabi et~al.(2018) used CLPsych 2015 shared task, containing Twitter posts tagged with depressed, control and PTSD as their training dataset. They also used Bell Let’s Talk~\cite{wikibelltalk} as the test dataset in their study. They used different Word embeddings such as Skip-Gram, CBOW, Optimised and Random and then fed them to various Deep Learning networks, three CNN-based and one RNN-based models
    for the evaluation phase.
    The CNNWithMax model, one of those three CNN-baseds, showed the best performance
    with an accuracy of
    87.957\% and precision of 87.435\%~\cite{orabi2018deep}. Also, in 2019, Arora et~al.~\cite{arora2019mining}, Ziwei et~al.~\cite{ziwei2019application}, Tao et~al.~\cite{tao2019twitter} and some more researchers studied
    depression on Twitter
    leveraging sentiment analysis techniques considering the
    sentences' polarities
    (positive, negative, or neutral). Besides, a recent study by Martins et~al.~\cite{martins2021identifying}, the NRC Emotion lexicon and NRC Affect Intensity lexicon were used along with ML and text mining techniques for their analysis. Table~\ref{tab:Dep_Idendification_Approaches} is a summary of different studies done with various approaches aiming at
    contributing to
    depression identification issues.

\end{itemize}
\subsubsection{Knowledge-Based Approaches}\label{KB_Section}


Knowledge base refers to a set of information and concepts, mainly in a specific field, being organised into a taxonomy, instances for each concept, and relationships among them~\cite{beheshti2019datasynapse, chai2013social}. Based on the literature, several studies have leveraged knowledge-based systems with this objective to develop various state-of-the-art automated systems. Knowledge-based summarisation systems~\cite{timofeyev2018building,ghodratnama2021intelligent}, Recommendation Systems~\cite{rosa2018knowledge,yakhchi2022convolutional,elahi2021recommender} and social streaming analytics~\cite{chai2013social} are some examples of
these kinds
of systems.

On the other hand,
a knowledge graph
is a kind of knowledge base, empowered by an inference motor~\cite{ehrlinger2016towards}. Knowledge graph is a complex net of various entities and their interrelations~\cite{paulheim2017knowledge} that could be related to a specific organisation or area.
Knowledge graphs can collect, extract, and integrate information about external resources~\cite{beheshti2020intelligent}.
These capabilities lead
to constructing an integrated knowledge-based system~\cite{ehrlinger2016towards}.

`Kosmix' is an example of knowledge-based systems, enabling various applications and helping with gaining
insight into a variety
of issues on social media. Kosmix is a framework for social data streams analytics and identifying the significant incidents that are becoming apparent. It is empowered by a scalable real-time data processing foundation such as RDBMs and Hadoop~\cite{chai2013social}.

The popularity of
using knowledge bases in different areas\footnote{https://www.ibm.com/cloud/learn/knowledge-graph} is illustrative of its power to help with finding new solutions and insights in
decision-making processes~\cite{beheshti2016business}.
Given the possibility of creating solutions for complex problems, KBs could be considered as beneficial tools for
problem solving
in a variety of fields, such as
mental health issues. In recent years, very few pioneer works conducted studies related to the using knowledge bases and the contextualised data-driven from them, for identifying mental and behavioural disorders.

\begin{figure}[t]
    \centering
    \includegraphics[width=0.65\textwidth]{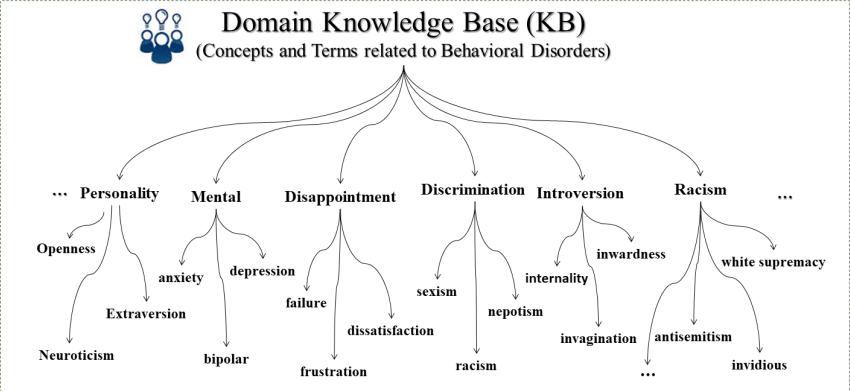}
    \caption{A segment of the domain-specific knowledge base for behaviour disorders~\cite{beheshti2020personality2vec}.}
    \label{fig:personality2vec_kb}
\end{figure}

Few recent studies tried to use knowledge-based approaches to formalise cognitive science knowledge related to behavioural disorders. For example,
Beheshti et~al.~\cite{beheshti2020personality2vec} provided
a new data analytics approach for analysing behavioural disorder patterns in social media platforms. They created a domain-specific KB based on the golden standards in personality and behavioural studies. Figure~\ref{fig:personality2vec_kb} demonstrate
an example snippet
of the domain-specific knowledge base for behaviour disorders, introduced by them.

They applied the KB for developing cognitive systems that intelligently contextualise and prepare raw social data for behavioural analytics. For this aim, they proposed
a word embedding approach based on patterns
to be implemented over each extracted feature
to
create a social behaviour graph model. As mentioned in their study, the proposed method could be applicable for identifying more specific mental disorders, such as depression and anxiety.
%
\subsubsection{Summary and paper Contribution}

Mental disorders were the leading sources of global health-related burden, with anxiety and depression problems accounting for the majority of cases.
In recent years, the global COVID-19 pandemic
has even worsened
the situation. People who are suffering from mental health conditions mostly visit a psychologist to receive supportive advice on their mental situations. In the clinical studies/approaches aiming at screening and identifying potential depressed patients and depressive symptoms, questionnaires
(e.g.,~\ref{fig:CES-D}~\cite{siwik2022depressive, lopez2021self} and ~\ref{fig:PHQ-9}~\cite{dadfar2021patient, costantini2021screening}) and also interviews are mainly used by psychologists~\cite{depressionmayo}.

Mental health disorders include a wide range of symptoms, risk factors, relationships, and effective interconnections. Most of those disorders~(e.g., depression) are not easily identifiable due to their complexity and need lots of time and effort to be identified~\cite{cramer2016major, fried2017mental}. As a result of complexity, mental disorders
will not be
easily automated or
translated into
the insights to help clinicians with their identification tasks.
It also causes
limitations for studies, aiming at using ML and automated analysis in this field. On the other hand, mental disorders such as anxiety and depression are complex subjective issues~\cite{cramer2016major, fried2017mental}. These complexities and subjectivity lead to the fact that having progress in mental disorder issues requires more time, study, and research.

On the other hand, as mentioned in section~\ref{Identification_Assessment_Approaches}, there are lots of evolving studies have been done, and various approaches
have been implemented
to help with the assessment,
screening, identification and prediction of depression.
On the other hand, those approaches have evolved to identify better and treat mental disorders.
For example,
as mentioned before, there are several versions of questionnaires developed during years aiming at progress in depression screening and identification. Although there are
valuable progress
and improvements in this field, there is still a huge need for further research and contributions to help with identification, prediction and treatment of mental disorders such as depression.

As our first contribution, we introduce a general-purpose Mental Disorder Knowledge Base~(MDKB).
Leveraging
knowledge-based approaches, the cognitive and psychological knowledge associated with mental health disorders are organised in the form
of a knowledge base,
providing a rich structure of relevant entities, semantics, and relationships among them. This knowledge base could help with decision-making processes by extracting insights
into the mental
conditions of potential patients.


\section{Method}
\label{Method}

In the previous Section, Section~2, we studied and analysed the various clinical and non-clinical approaches to identifying mental health disorders.
We highlighted that state-of-the-art research in identifying
Mental Disorder~(MD)
patterns from textual data, uses hand-labelled training sets, especially when a domain expert's knowledge is required to analyse various symptoms. This task could be time-consuming and expensive.
To address this challenge, in this Section, we present a novel approach to facilitate mining of mental disorder patterns from textual data.
We leverage the domain knowledge and expertise in cognitive science to build a domain-specific Knowledge Base (KB) for the mental health disorder concepts and patterns.
We present a weaker form of supervision by facilitating the generating of training data from a domain-specific KB.

\section{Constructing A Domain-Specific Knowledge Base}
\label{MDKB_construction}

\begin{figure}[t]
    \centering
    \includegraphics[width=0.9
    \textwidth]{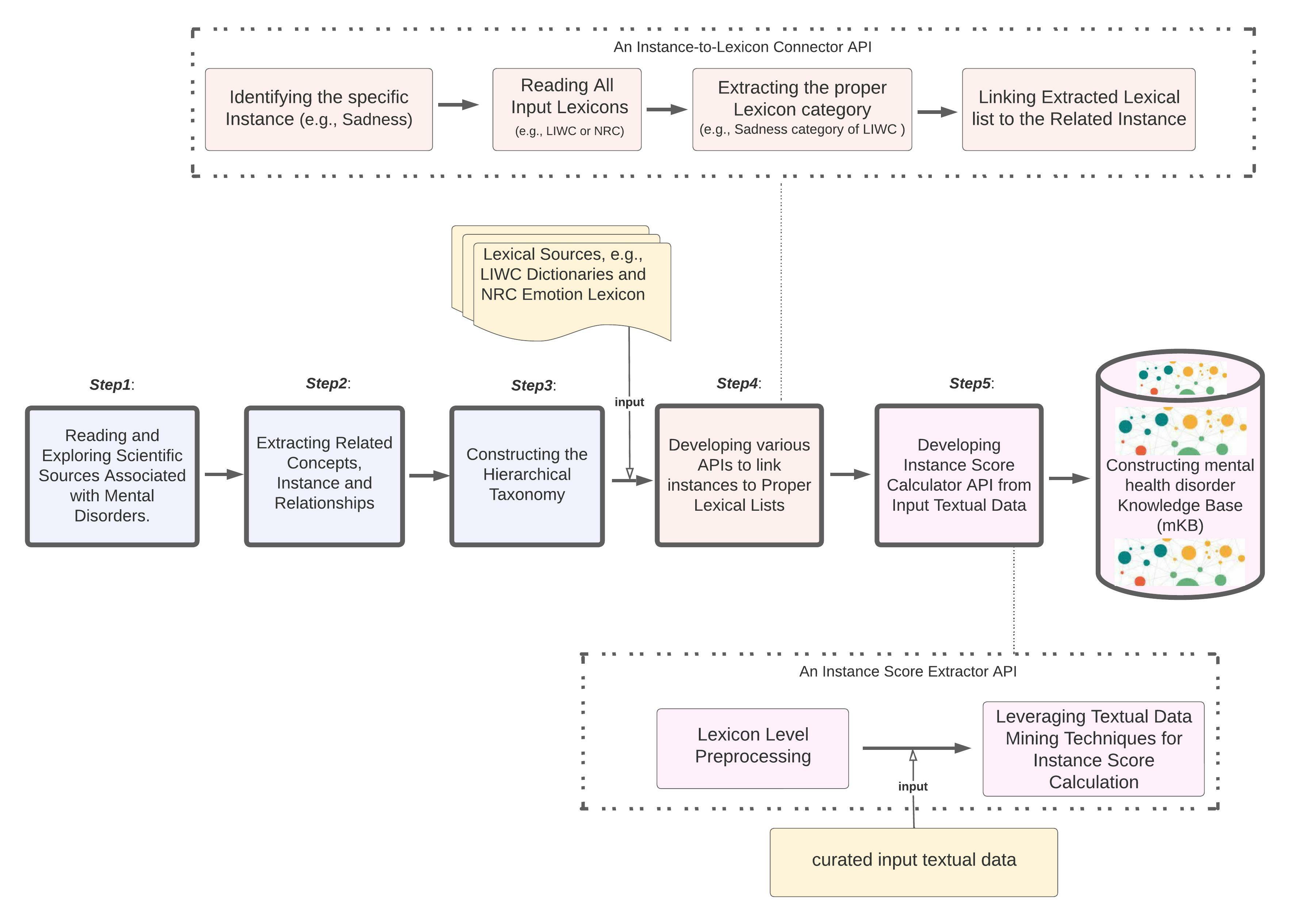}
    \caption{The five-step process of constructing mental disorder knowledge base.}
    \label{MDKB_Construction}
\end{figure}

In this section, we explain our methodology to leverage the domain knowledge and expertise in cognitive science to build a domain-specific Knowledge Base (KB) for the mental health disorder concepts and patterns.
As illustrated in Figure~\ref{MDKB_Construction}, construction of the proposed
mental health disorder Knowledge Base (mKB)
consists of five steps:
%
(i)~exploring scientific sources;
(ii)~extracting related concepts, instances and relation-
ships;
(iii)~constructing the hierarchical taxonomy;
(iv)~developing instance-to-lexicon connector APIs; and
(v)~developing instance score calculator APIs from input
textual data.

\subsection{Step 1: Exploring Scientific Sources}

In the first step,
a set of golden standards in cognitive computing
that are related to mental disorders are
explored.
There are
several best practices, research, online sources, guidelines, and manuals associated with mental disorder identification, helping with constructing
the mKB.
For example, some of the
golden standards and best practices in this field are `Diagnostic and Statistical Manual of Mental Disorder'~\cite{edition2013diagnostic} and `Clinical Practice Guidelines and Principles of Care for People with Dementia', developed by the American psychiatric association and the Australian Government National Health and Medical Research Council (NHMRC)~\cite{dementia}, respectively.

In addition, `ICD-11', a book by World Health Organisation\footnote{https://www.nice.org.uk/guidance/ng54} and `guidance on assessment, diagnosis, and management of dementia', published together by the British Psychological Society and Royal College of Psychiatrists~\cite{coooper2016mental} are among those sources. Also, `panic disorder, social anxiety disorder, and generalised anxiety disorder guide'~\cite{andrews2018royal,abu2019towards} contains guidelines for Clinical measures developed by `Royal Australian and New Zealand College of Psychiatrists'.
These sources could be used for constructing the
proposed mKB.

\subsection{Step 2: Extracting Concepts, Instances and Relationships}

In the second phase of constructing
the mKB, after
exploring previous studies and golden standards such as those mentioned in Section~\ref{Background_Section}, various 
mental health related
Concepts, instances, and relationships
need to be extracted.
For example,
in designing the mKB, we considered
`Anxiety Disorders', 'Depressive Disorders', `Neurocognitive Disorders', `Paraphilic Disorders' and `Bipolar and Related Disorders'
as instances of the concept `Mental Disorders'.
Also, anxiety disorders, distinguished by a special way of thinking and behaviours~\cite{andrews2018royal},
contain instances
such as `panic disorder', `social anxiety disorder', `agoraphobia', and `generalised anxiety disorder'.

Each of these instances would probably consist of some concepts, such as `Risk Factors', `Symptoms', and `Supportive Symptoms'. As an example, `Negative encounters' could be a risk factor for social anxiety disorder and are very likely to occur in children who are ridiculed, abused, and neglected~\cite{socialanxiety}. Therefore, negative encounters
could be considered as instances of
the social anxiety disorder risk factors. On the other hand being bullied, rejected, and also ridiculed are considered to be instances of negative encounters. Social anxiety disorder has symptoms such as fear or anxiety attacks, considered to be its instances.~\cite{andrews2018royal, edition2013diagnostic}. The rest of the concepts, instances, and relationships in the mKB are extracted in the same way as explained for anxiety disorders.

\subsection{Step 3: Constructing the Hierarchical Taxonomy}

In this step, all of the extracted
mental health disorders related concepts
and instances are organised into a taxonomy, being a  hierarchical structure for classifying those concepts and instances.
The mental health disorder taxonomy
contains six main
levels. The main concept~(i.e., `Mental Disorders') is placed on the top level. The second level contains various instances of mental disorders~(e.g., `Anxiety Disorders') each of which represents a specific group of diseases. For example, `panic disorder', `social anxiety disorder', `agoraphobia', and `generalised anxiety disorder' are contained in a group that `Anxiety Disorders' is its representative. All of the instances in this level are connected to the previous level concept~(i.e., `Mental disorders') and also to the next level~(i.e., third level) instances which are all single disorders contained in a group of related disorders as mentioned before.

\begin{landscape}
\begin{figure}[h]
    \centering
    \includegraphics[scale=0.4]{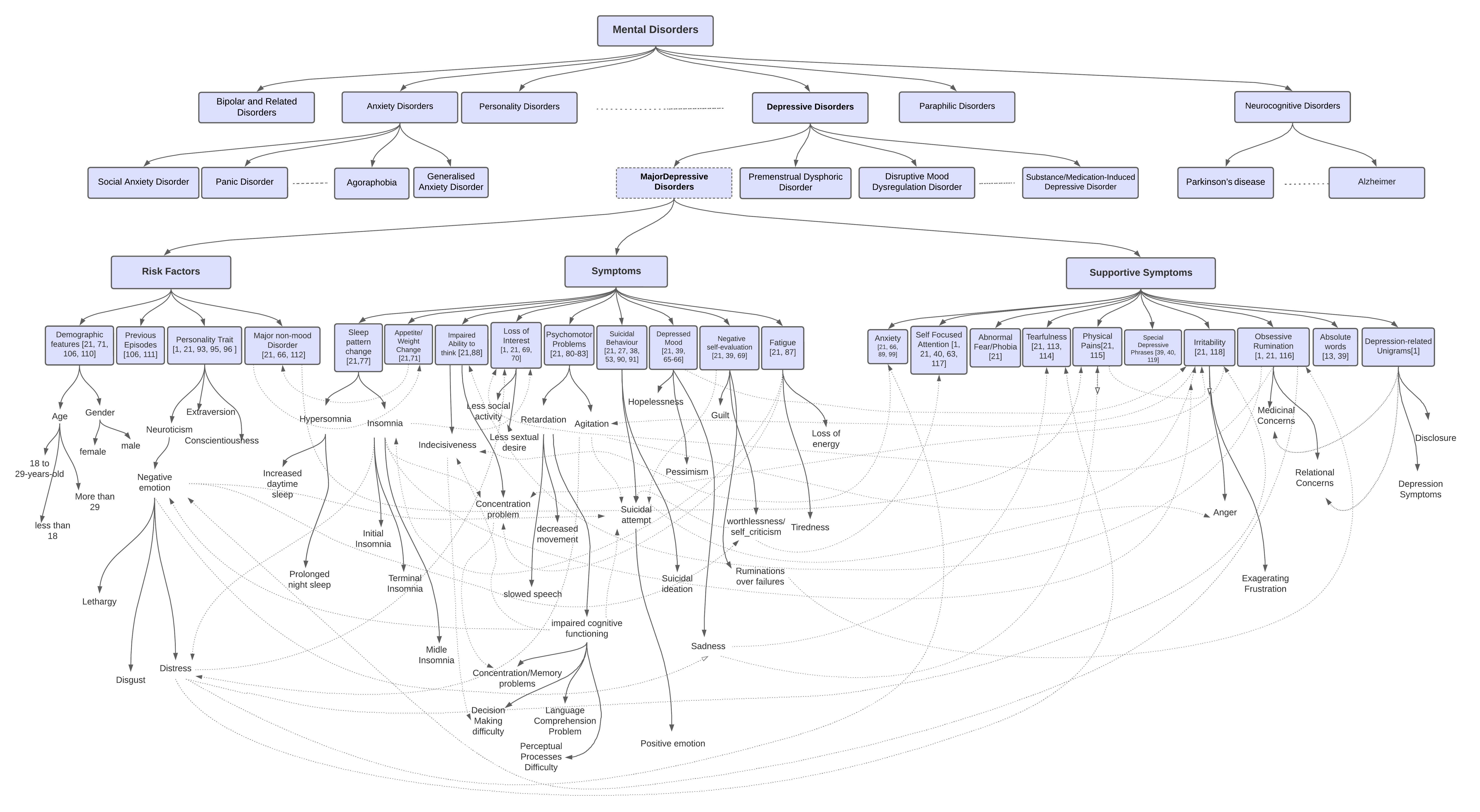}
    \caption{A snapshot of the mental health disorder Knowledge Base focuses on depression, being one of the main causes of disability globally~\cite{de2013predicting, lemoult2019depression}. It contains different concepts related to each mental disorder, instances of each concept, and the relationships among them. Vertical solid connections are mainly indicative of relationships between concepts and their instances~(e.g., impaired cognitive functioning is an instance of retardation~\cite{buyukdura2011psychomotor}), while horizontal dotted connections are illustrating the different kinds of probable relationships~(e.g., impaired cognitive functioning could cause negative emotion)~\cite{lemoult2019depression}.}
    \label{fig:Depression_KB}
\end{figure}
\end{landscape}

Each specific instance in the third level is linked to a set of related concepts
in the fourth level. These concepts in the fourth level are `Risk Factors', `Symptoms', and `Supportive Symptoms'. Each concept in the fourth level is connected to its instances in the fifth level and also to the probable sub- and sub-sub- instances in the sixth level and afterwards. All the above-mentioned
connections~(i.e., from up to down)
are shown by solid arrows. However, it is possible that some Concepts and instances would be horizontally connected to each other. Those horizontal connections~(i.e., relationships) are shown by dotted arrows.
%
Figure~\ref{fig:Depression_KB} is a snapshot of the mental health disorder Knowledge Base focuses on depression, being one of the main causes of disability globally~\cite{de2013predicting, lemoult2019depression}. It contains different concepts related to each mental disorder, instances of each concept, and the relationships among them. Vertical solid connections are mainly indicative of relationships between concepts and their instances~(e.g., impaired cognitive functioning is an instance of retardation~\cite{buyukdura2011psychomotor}), while horizontal dotted connections are illustrating the different kinds of probable relationships~(e.g., impaired cognitive functioning could cause negative emotion)~\cite{lemoult2019depression}.

\subsection{Step 4: Developing Instance-to-Lexicon Connector APIs}

Kinds of words~(e.g., self-focused words, hate speech, positive or negative emotion, death-related, anger), sentence sentiments, and natural language that are used in the daily communications are reflections of individuals' real-life manners, thoughts, cognitive processes, behaviours, and mental health status~\cite{glenn2020can, serani2011living, al2018absolute, de2013predicting, tausczik2010psychological, schmidt2019survey, troop2013expressive, de2013major}. According to
numerous studies,
there is a clear link between a set of words and the presence of distinctive content~(e.g., profane~\cite{rezvani2020linking}, self-criticism~\cite{troop2013expressive}, negative emotion~\cite{de2013major}, suicidal attempt/ideation~\cite{fernandes2018identifying}) in textual documents.

As a result,
we proposed to use
lexical resources for depression identification through text analyses. Hence, prior to this step,
various
lexical resources are fed into
the mKB-related
taxonomy, aiming at turning that taxonomy into a
lexical-driven knowledge base,
which facilitates textual data mining.
The goal is to use a collection of related words that can be used for extracting the context-specific features~(i.e., linguistic or emotional features) from the text.
Some examples of valuable lexicons are LIWC\footnote{http://www.liwc.net/}(i.e., psycholinguistic lexicon), ANEW~(Affective Norms for English Words)~\cite{bradley1999affective, nielsen2011new}, NRC~\cite{mohammad2017word, mohammad2013crowdsourcing}, Warriner Lexicon~\cite{warriner2013norms}, and MPQA~\cite{wiebe2005annotating}. These lexicons contain several categories, related to a specific emotion, behaviour, affect, or linguist style associated with an instance in the MD taxonomy.

For example, LIWC contains the `sadness' category, consisting of a list of around 1,300 words related to sadness. On the other hand, under the `Depressed mood' category of the MD-related taxonomy, there is an instance named `sadness'. In this step, these associated items will be linked together by leveraging several APIs. A unique API is developed for each of the instances in the fifth level or their sub-/sub-sub instances in the next levels, and embedded into
the mKB.
Each of these APIs
is a Machine Learning~(ML)
algorithm that links the corresponding instance~(i.e., included in the MD-related taxonomy) to one of the proper input lexicons, as well as the proper category of that lexicon. This step helps with constructing an MD-related taxonomy enriched with thousands of words~\cite{amouzgar2018isheets}.

\subsection{Step 5: Developing Instance Score Calculator APIs from Input Textual Data}
\label{Step5_section1}

In this step, several APIs are developed, each of which is associated with one of the instances in the MD-related taxonomy. APIs are ML algorithms, that calculate the feature scores~(e.g., sadness score), corresponding to each
instance~(e.g., sadness),
via analysing input textual data. calculated feature scores act as fundamental parts of our
proposed mKB,
which enables
automated knowledge-based analysis. One of the automated tasks, empowered by using this knowledge-based analysis is the identification of probable individuals with mental disorders. For example, through this research, a classifier is developed~(i.e., explained in Section~\ref{Developing-Classifier}), that enables identifying potential depressed individuals, automatically. The main tasks, enabled by the instance score calculator APIs are as follows:

\begin{itemize}
    \item\textbf{Lexicon Level Preprocessing:} Prior to any score calculations, all the words included in each instance-related lexicons need to be preprocessed. The preprocessing consists of removing unwanted and useless symbols, and punctuation and also stemming all the words included in each lexicon. Stemming is the technique of reducing words to their word stem, base, or root form\footnote{https://en.wikipedia.org/wiki/Stemming}. Since the lexicon words will be used for calculating similarity scores between each lexicon category and the input text, stemming could be the
    most significant and effective technique in this stage.
    \item\textbf{Linking the Curated Input Textual Data:} As mentioned before, each instance in the taxonomy is connected to the proper lexicon. The instance score calculator API enables linking the curated input textual data to the related instance and lexicon. Data curation is an automated process that prepares the raw input data prior to further data analysis~\cite{beheshti2019datasynapse}. In Section~\ref{curation-step},
    the data curation process
    is described in more detail.
    \item\textbf{Leveraging Textual Data Mining Techniques for Instance Score Calculation:} After cleaning and preparing the raw data, different features are extracted
    from the cleaned data.
    There are two types of features that could be extracted from the input cleaned/prepared data, namely text-related features and user-related features. Sentiment polarity~(i.e., positive, negative, or neutral feeling
    conveyed
    through a text) and personality trait~(e.g., being neurotic or extravert) analysis are examples of text-related and user-related features, respectively. Afterwards, our data is enriched by different techniques such as stemming. Finally, after data cleaning, preparation, feature extraction, and feature enrichment phases, the output~(i.e., featured data) is ready for further analysis and score calculation process~\cite{beheshti2017systematic}.

    In our proposed method,
    different scores
    such as cosine similarity scores, suicidal behaviour scores, and personality trait scores are calculated. Cosine similarity scores help us to identify the rate of similarity between the input textual data~(i.e., written by a potential depressed person) and the instance-related lexicons. Those scores are expressive of how frequently an MD
    instance-related words~(e.g., sadness-related words) are used by a specific individual. Consequently, they could be indicative of a specific MD-related instance~(e.g., sadness feelings) that the writer conveys and shows through his written text.

    Cosine similarity scores are calculated by the instance calculator APIs. They are computed using different textual data mining techniques, such as NLP techniques, consisting of `Term Frequency–Inverse Document Frequency'~(TF-IDF) and `Cosine Similarity'. TF-IDF is a text mining technique for determining the importance of a word in a collection or corpus of documents\footnote{https://en.wikipedia.org/wiki/Tf\%E2\%80\%93idf}. Each word in text mining is given a unique coordinate, and a document is represented by a vector representing the number of times each word appears in the document. Also, cosine similarity is a useful metric for determining how similar two texts are likely to be in terms of content, regardless of their length\footnote{https://en.wikipedia.org/wiki/Cosine\_similarity}.

    On the other hand, motivated by a recent study~\cite{majumder2017deep}, the probability of a special kind of personality trait~(e.g., neuroticism) could be extracted. This probability is calculated by a specific API, connected to the personality trait instance
    on the mKB.
    This API consists of a sophisticated algorithm, using a 7-layer convolutional neural network and several NLP techniques such as word2vec embedding. In addition, Suicidal behaviour~(e.g., suicidal attempts or ideation) scores are calculated using a special score calculator API.

    In addition, there could be other APIs that don't need text mining techniques for score calculation. For example, the `Gender' or `Age' instance-related APIs would create simple numerical scores. For example, `Gender' API's scores could be `0' or `1', indicative of the male or female users. Also, `Age' API could give some scores based on different pre-defined age ranges. Algorithm~1 demonstrate the process of instance score calculation for those APIs that apply textual data mining techniques for this aim.

    \begin{algorithm}
     \KwData{Curated Input Textual Data}
     \KwResult{calculated Instance-related Scores}
     The mKB Instances;\\
     Linked Instance-related lexicons;\\
      \For{Each Instance in the mKB} {
        Reading the Curated Input Textual Data;\\
        Reading the linked lexicon;\\
        Applying proper text mining techniques~(e.g., Cosine Similarity Calculation);\\

     }
     \caption{The process of instance score calculation in API's that work based on textual data mining techniques.}
    \end{algorithm}
\end{itemize}

Finally, developed APIs are linked to corresponding instances, and embedded in the taxonomy. After embedding the APIs into the taxonomy, it technically turns into a knowledge base. Through this KB several mental disorder scores would be calculated. These scores could be shown as an extracted knowledge
from mKB
aiming to give some insight and alarms to the analysts or psychologists. This knowledge besides analysts' experiences could probably be informative, helping them make more accurate identifications.

\section{Developing a Weakly Supervised Classifier for Target Mental Disorders Using Domain-specific KB}
\label{Developing-Classifier}

ML algorithms have notable real-world applications. Most of these applications are enabled by leveraging deep learning models and existence of various open-source ML platforms such as TensorFlow and PyTorch, as well as a large number of modern models. However, large hand-labelled training datasets are required for these models to perform well~\cite{ratner2019weak}.

In recent studies, hand-labelled training data sets have been employed to find mental disease patterns from textual data. Benefiting from experts and psychologists knowledge to label a dataset with presence or absence of a disorder could be an expensive and time consuming task. Besides, evolving nature of knowledge, makes it necessary to access re-labelling mechanisms for more accurate data analysis tasks~\cite{ratner2019weak}. In cognitive science area, which includes high levels of subjective knowledge, having access to labelled data for analytical tasks is even more crucial~\cite{beheshti2020towards}.
Weak supervision is a learning approach in which a higher level source of supervision can be used to generate a larger training sets~\cite{ratner2019weak}. In this section, we leverage the mental health domain-specific Knowledge Base (mKB), discussed in previous section, to propose a weakly supervised mental disorder classifier.
Using the cognitive and psychological knowledge that is embedded in the mKB acts as a kind of domain experts knowledge to be used as a labelling asset, contributing to the creation of more labelled training data sets for cognitive analytics.

Our approach will enable the mKB to be the main source of supervision. The mKB, is a general KB, consists of several instances of mental disorders and their related concepts and instances. Our proposed approach is applicable at a time to one of the mKB's disorders~(e.g., depression), which is called Target Mental Disorder~(TMD). The mKB, is empowered with several APIs, which are calculating TMD-related instance scores. These scores are valuable features, associated with specific cognitive, behavioural, and mental conditions, which enable the process of labelling training data sets for any other studies related to mental disorders.

\begin{figure}[h]
    \centering
    \includegraphics[angle=0, scale=0.45]{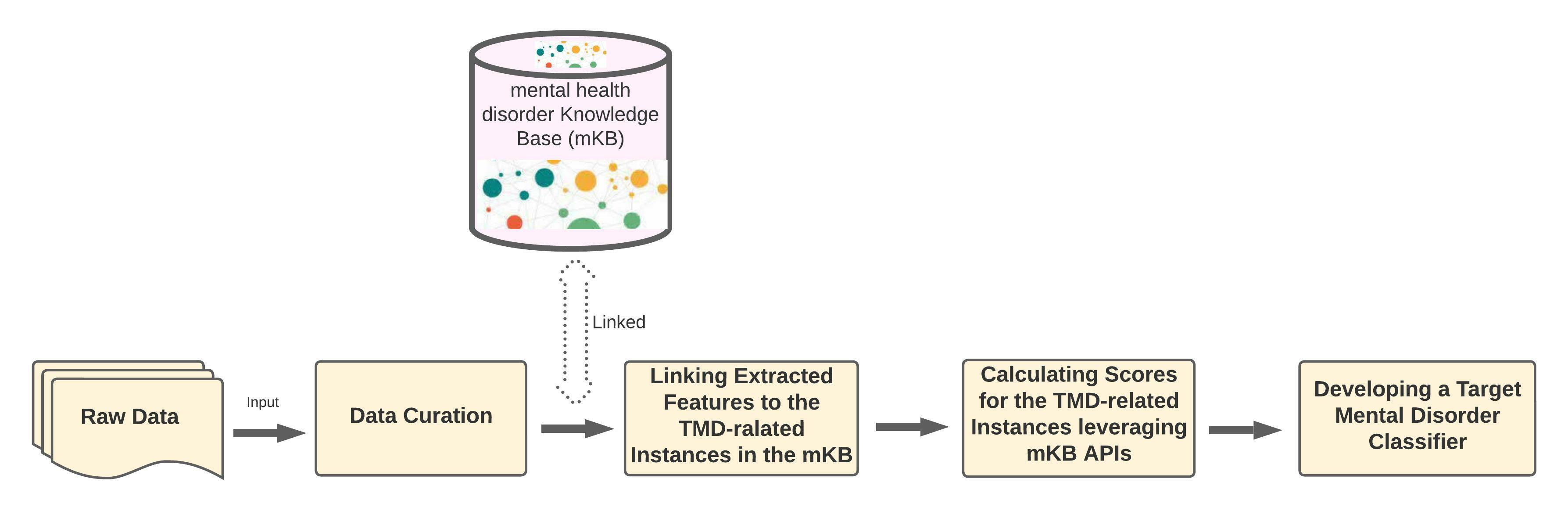}
    \caption{The proposed pipeline for developing a weakly supervised mental health disorder classifier, leveraging calculated scores for Instances related to a Target Mental Disorder~(TMD), generated by mKB.}
    \label{fig:Classifier_development}
\end{figure}

Figure~\ref{fig:Classifier_development}
illustrates the
proposed pipeline for developing a target mental disorder classifier. There are several stages included in developing the proposed pipeline, as follows:
(i)~`Data Curation';
(ii)~`Linking Extracted Features to TMD-ralated Instances in mKB';
(iii)~`calculating Scores for TMD-related Instances leveraging mKB APIs'; and
(iv)~`Developing a Target Mental Disorder Classifier'.
Our proposed approach offers an extensible framework by adopting a service oriented architecture~\cite{ProcessAtlas} to enable new analytical features to be plugged into the system in an easy way.
For example, if we find it effective to add other features~(e.g., `swear words', `hate speech' or `profane words' scores) to our model,
then the extensible architecture will facilitate this for the analyst.

\begin{figure}
    \centering
    \includegraphics[scale = 0.5]{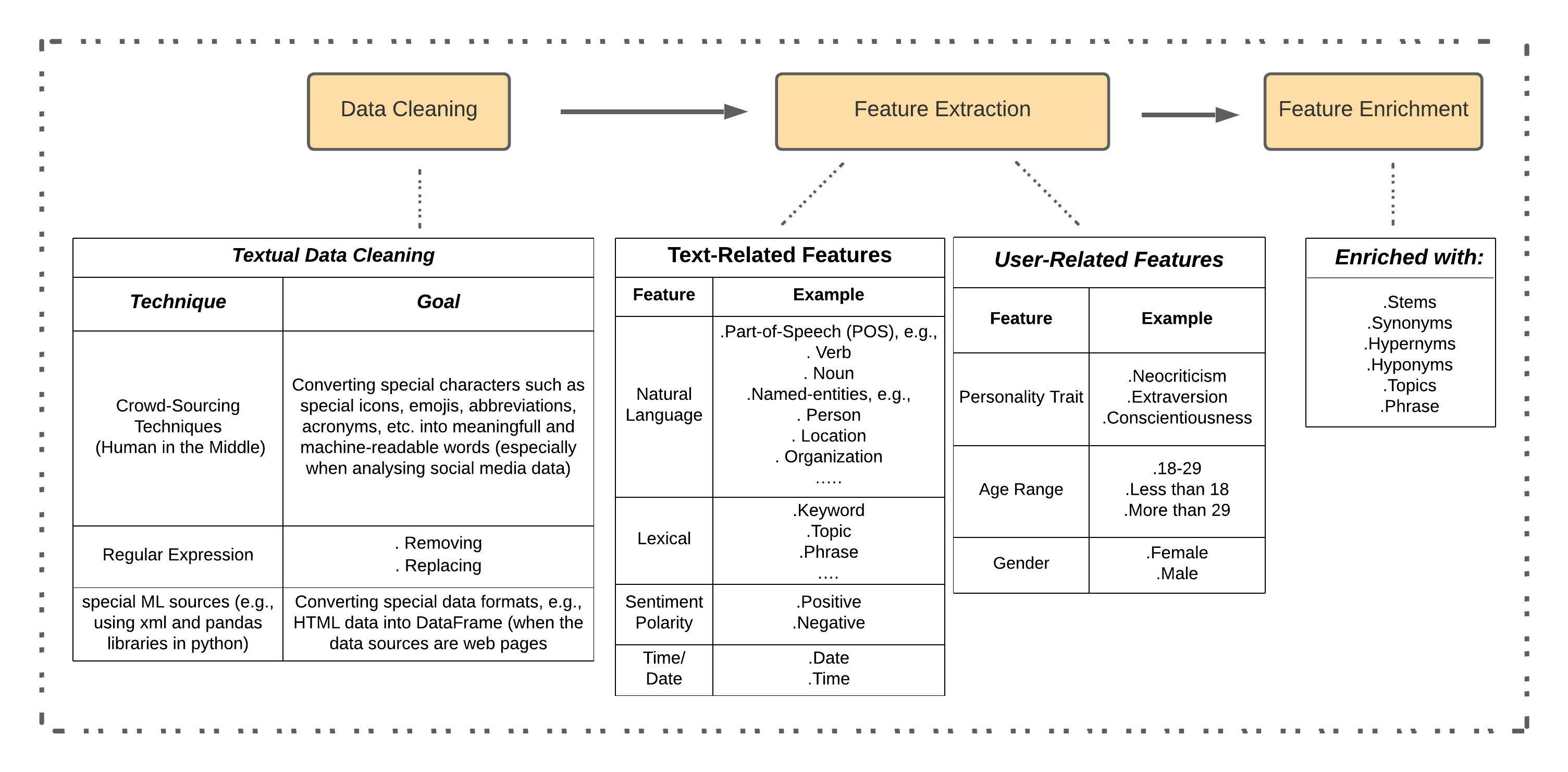}
    \caption{Data curation process motivated by Beheshti et~al. study~\cite{beheshti2019datasynapse}.}
    \label{data_curation}
\end{figure}

\subsection{Data Curation}
\label{curation-step}

Data curation defines as a process that turns the raw data into contextualised data and knowledge~\cite{beheshti2019datasynapse}. This process helps the ML algorithms run more efficiently. Motivated by two recent studies~\cite{beheshti2018corekg,beheshti2019datasynapse}, as Figure~\ref{data_curation} demonstrates, the raw textual data needs to be prepared and curated before any further analysis. The curation process contains 3 phases, namely (i)~Data Cleaning; (i)~Feature Extraction; and (iii)~Feature Enrichment, which are explained as follows.

\begin{itemize}
    \item\textbf{Data Cleaning: }There are different techniques and tools that could be used for cleaning textual data. For example, with the help of some ML packages and their corresponding methods(e.g., XML, Pandas, and RegEx~(i.e., regular expressions)) useless data~(e.g., punctuation, numbers, symbols, etc) could be removed or some special parts of raw data~(e.g., acronyms such as PSTD) could be replace with the proper real words~(e.g, Post Traumatic Stress Disorder
)). Also, special noisy data formats~(e.g., HTML and web-related data) could be turned into much easy-to-work formats~(e.g., DataFrame). Besides, crowd-sourcing is a modern approach for cleaning social data. In this approach, the knowledge of the crowd~(e.g., Amazon Mechanical Turks) is used for some correction tasks~(e.g., spell- or abbreviation-checking).

    \item\textbf{Feature Extraction: }There are two types of features that could be extracted from the clean data, namely `Text-Related Features' and `User-Related Features'. The text-related type consists of `Natural Language', `Lexical', and `Time/Date' features . The user-related type consists of the user's `Personality Trait', `Age Range', and `Gender'. Some of these features are explained as follows.
    \begin{itemize}
        \item\textbf{Natural Language Feature} refers to entities that could be retrieved from Natural Language and speech through ML analysis. It includes Part-of-Speech, named entities, etc. Verbs~(e.g., developed, caused, felt, experienced ), Nouns~(e.g., psychiatrist, depression, anxiety, acid Reflux, Acticlate, Zoloft ), and adjectives~(e.g., nauseous, busy, terrible, bothering) are some examples of the Part-of-Speech. Person names~(e.g., Jean Piaget and Albert Bandura), Organisations' names ~(e.g., Royal Prince Alfred), locations~(e.g., Sydney) are also some instances of named entities~\cite{hosseinpoor2020proposing}.
        \item\textbf{Lexical Feature} includes examples such as Keyword~(e.g., hypochondria, depressive episode, depression), Topic~(e.g., Daily Acid Reflux\footnote{A digestive disease in which stomach acid or bile irritates the food pipe lining.}), Phrase~(e.g., 'with my psychiatrist's help'), Abbreviation~(GAD and IBS), and Spelling Errors.
        \item\textbf{Time/Date Feature} refers to the time~(14:14:54) or date(2017/02/21), when a text is created. This feature could help with getting insight into the probable sleep time or when a user is mainly active in social media. It could be a valuable feature in the time series analysis.
        \item\textbf{Personality Trait Feature} consists of different personality traits~(e.g., neuroticism, extraversion, openness, etc). It is one of the user-related features, extractable via different modern textual analysis techniques~\cite{majumder2017deep}. A personality trait could contribute to suffering from several mental conditions~\cite{kendler2007sources, kendler2004interrelationship, kendler2006personality}, and could be an important feature to be considered in the cognitive analysis.
    \end{itemize}

    \item\textbf{Feature Enrichment} relates to utilising relevant sources and services~(e.g., WordNet\footnote{wordnet.princeton.edu/} and STANDS4\footnote{abbreviations.com/abbr api.php}) to identify synonyms and stems for a keyword that has been extracted. For example, 'traumatic' as a keyword could be enriched with its synonyms such as 'disturbing', 'shocking', 'distressing' and 'upsetting'. IBS and GAD are two keywords which are extracted from the text. IBS and GAD are two acronyms, that are equivalent to two types of disorders, namely `Inflammatory Bowel Disease' and `Generalized Anxiety Disorder'. Identifying these equivalents relationships leads to creating more enriched data to help with accurate data analysis tasks.
\end{itemize}

\subsection{Linking Extracted Features to the TMD-related Instances in the mKB}
\label{linking-step}

Each mental disorder included in the mKB~(e.g., depression) contains several instances~(e.g., `Sadness', `Suicidal Behaviours',`Gender'). To link these instances to the curated data and extracted features, each instance is provided with an empty list. Then, extracted features~(e.g., stemmed keywords or phrases) are added to the corresponding instance list. The kind of added features to the list is dependent on the type of the related instance. For example,  extracted gender(i.e., male or female) related to the writer of a text is
added to the empty list provided for the instance of `Gender' in the mKB.

Besides, for `Sadness' instance, for example, we need
extracted features
to be added to the list. On the other hand, there are some
instances such as `Suicidal behaviour' that need the
whole raw
text~(i.e., written by each user), instead of extracted features, to be added to the corresponding list,
aiming at counting the number of suicidal phrases via corresponding API. Algorithm~2 illustrates the process of linking extracted features to the corresponding instances in the mKB.

\begin{algorithm}
 \KwData{Curated input textual data}
 \KwResult{List of features that are linked to the instances of the mKB}
 Specify the TMD;\\
 Extract Features from textual data;\\
 TMD\_ralated\_Instances = Set up and Array for TMD-ralated Instances in the mKB; \%(e.g., `Negative Feeling', `Anxiety', `Sadness', etc.) \\
 \For{Each TMD\_ralated\_Instances } {
    Generate an empty list;\\
        {
 \For{Each feature in Extracted-Feature} {
    Add the feature to the corresponding instance-related list;\\
    Link Extracted Features to the instances in mKB;
    }
   }
  }
 \caption{Linking Extracted Features from input textual data to the TMD-ralated Instances in the mKB.}
\end{algorithm}

\subsection{Calculating Scores for the TMD-related Instances Leveraging mKB APIs}
\label{score-calculation-step}

The mKB is empowered by several APIs for calculating the instance scores. Each of those APIs has a separate functionality. Some of them~(e.g., the `Suicidal Behaviour' or `Special depressive phrases' API) create scores that are indicative of the counted numbers of special phrases that are use in a cleaned text. On the other hand, some of them~(e.g., `personality Trait' API) create a score that is the probability of having a special personality trait~(e.g., neuroticism) by analysing a cleaned text.

The linked extracted features~(i.e., from Section~\ref{linking-step}) are the input for corresponding instance-related APIs. This input enables the score calculation processes for each of the TMD-related Instances. For example, a list of extracted stemmed tokens is linked to the `Sadness' API. This API calculates the cosine similarities between the linked list and the related lexical source, being attached to the instance. The calculated scores act as input features for developing a learning algorithm~(i.e., a classifier), and are fed into the next step. Algorithm~3 demonstrates the process of calculating scores for the TMD-related instances, leveraging mKB APIs.

\begin{algorithm}
 \KwData{Linked Features~(i.e., the Algorithm~2 output)}
 \KwResult{Scores Related to Instances}
 Specify the TMD;\\
  \For{Each TMD-related Instance in mKB} {
    Instance-related Score = Embedded-score-calculator-APIs(linked features);\\
    \%Embedded-score-calculator-APIs() is a function for calculating scores\\
    \%functions mainly use techniques related to TFIDF or Cosine Similarty calculation
 }
 \caption{Calculating Scores for TMD-related Instances Leveraging mKB APIs.}
\end{algorithm}

\subsection{Developing a Target Mental Disorder Classifier}

In this step, the calculated scores for each TMD-related instance act as input features of the TMD-related classifier. several ML algorithms such as Random Forest, Logistic Regression, SVM and XGbooster could be deployed to develop this classifier. After feeding related scores, a binary learning algorithm is developed to classify whether an input text is associated with a target mental health disorder or not.

For example, considering depression as a TMD, the corresponding classifier would take a textual document and give it the labels of `0' or `1' for `not depressed' or `depressed' tags, respectively. In the next Section, we will discus how we create a depression-related classifier with a high accuracy, using ensemble techniques. Algorithm~4 illustrates the process of developing a TMD-related classifier.

\begin{algorithm}
 \KwData{Calculated Instance-related Scores}
 \KwResult{Developed TMD-related Classifier}
 Specify the TMD;\\
  \For{Each TMD-related Instance} {
    Generate
    Instance score using related API;\\
    Integrate all the scores as a feature array;\\
    Deploy a Proper ML classifier using the feature array;\\
     }
 \caption{Process of Developing a TMD-related Classifier.}
\end{algorithm}


\section{Evaluation and Experiment}
\label{Evaluation}

In this Section, the outcomes derived from the proposed method in Section~\ref{Method} are explained and evaluated. We will demonstrate how this model can help analysts and domain experts to get insight into the mental health status of their target population. First, a depression-related motivating scenario is described, clarifying the importance of this research.
Afterwards,
we present
the experimental setup, the dataset, experimental results, and evaluation of our method.

\subsection{Motivating Scenario}

The mental health of individuals and communities is a pressing challenge in the world, nowadays.
Since COVID-19\footnote{https://en.wikipedia.org/wiki/COVID-19} pandemic outbreak
in 2019, most governments have been preoccupied with handling and combating the epidemic. After the global success in vaccine development, a key concern for most governments is to harness the impacts of years of virus exposure, including the associated psychological problems in the community.
Based on a recent study, after being diagnosed with Covid, roughly one out of every five people develops a mental disorder~\cite{taquet2021bidirectional}. Data from the United States and Australia show elevated rates of depression throughout the epidemic; It is estimated that during the outbreak, depression level~(i.e., 25\%) is seven times higher than pre-pandemic levels worldwide~(i.e., less than 4\%)~\cite{covidinaustrali}. Hence, it could be a critical issue for governments to monitor, identify and take proper actions in this regard.


Due to the importance of the issue, as a motivating scenario, we focus on depression.
We use the framework proposed in Section~3
to highlight how our method significantly assists
e-safety\footnote{https://www.esafety.gov.au/} community
to monitor and identify
the trend for the emergence of depression-related symptoms, during special periods of time such as the Covid-19 period.
We leverage the
domain specific KB
(that discussed in Section~3)
and expertise in cognitive science to build a domain-specific Knowledge Base (KB) for depressive patterns. This KB contains a set of depression-related concepts organised into a taxonomy, instances for each concept, and relationships among them.
Then, we use the proposed
weakly supervised learning approach (Section~3) by facilitating the generating of labelled training data from depression-related KB.
To evaluate our approach, we adopt a typical scenario for analysing social media to identify major depressive disorder symptoms from the textual content generated by social users.

\subsection{Experimental Setup}

\subsubsection{Experiment Environment}

We use Jupyter notebook from Anaconda~3 environment, Python version '3.9.0', Pandas~('1.3.4'), NLTK~('3.5'), Sklearn~('1.0.1'), XGBoost~('1.5.0') for performing our NLP, textual data analysis, and developing depression-related classifier.

\subsubsection{Dataset}
\label{data_set}

People's
ideas and feelings can be viewed through social media platforms such as Twitter. New studies have centred on the investigation and mining of such data in a range of fields, including financial systems, politics, and healthcare~\cite{de2013major}. On the other hand, social media is a rich source of textual data, enabling researchers and data scientists analyse cognitive processes and behavioural patterns of the social media users~\cite{de2013predicting,schiliro2020cognitive}. Hence, as we are interested in identifying potential depressed individuals~(i.e., part of our motivating scenario) through textual data mining,
we use a real-world dataset\footnote{https://github.com/BigMiners/eRisk2018} from Reddit\footnote{https://en.wikipedia.org/wiki/Reddit}, i.e., a social news aggregation, web content rating, and discussion Website.
%
The dataset contains
820 users, each tagged with `1'~(i.e., `depressed') or `0'~(i.e., )
`not depressed' labels.
From a total of 820 users, 79 were tagged as depressed and the rest~(i.e., 741 users) were tagged as not depressed. The raw data is in the form of XML files and it is organised into ten chunk's. Each chunk contains 820 folders~(each folder for one user), and each folder consists of parts of the posts related to one of the 820 users. Considering all of the posts related to each user~(contained in all 10 chunks), the average number of posts for each user is 664 numbers.

\subsection{Experimental Results}\label{Experimental_Results}

In this part, we describe the experimental results derived from our method. First, we explain the construction of depression-related KB. Then, we explain the process of developing a weakly supervised classifier for depression, using the results derived from the constructed depression-related KB.

\subsubsection{Constructing a Depression-related Knowledge Base}

\begin{figure}
        \centering
        \includegraphics[scale = 0.45]{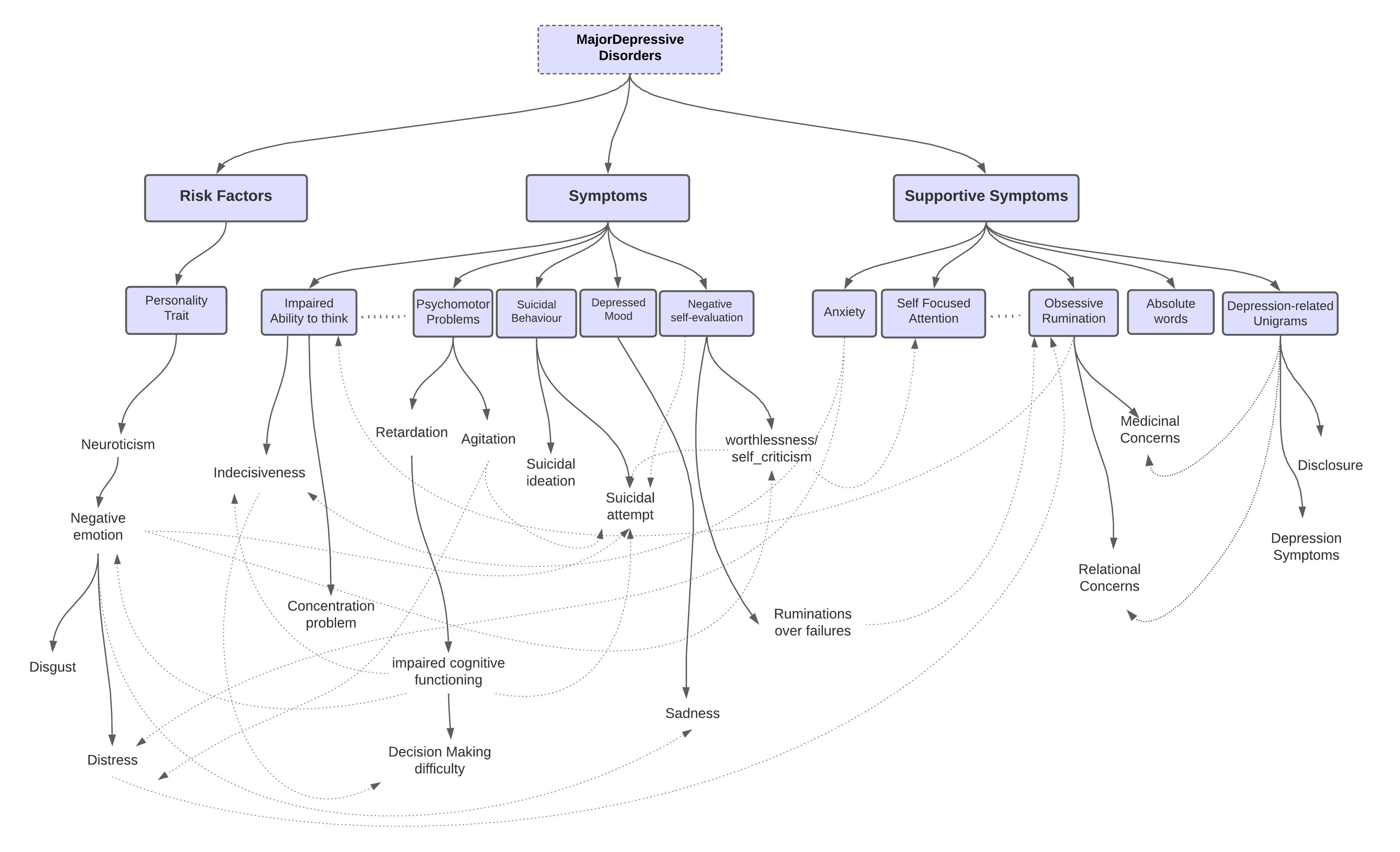}
        \caption{A Snapshot of depression KB, containing some of concepts and instances used in evaluation part.}
        \label{fig:Depression_KB_eval}
    \end{figure}

As mentioned before, a knowledge base consists of a taxonomy, several concepts, sub-concepts, instances, and the relationships between them. Figure~\ref{fig:Depression_KB_eval}
illustrates a
snapshot of the depression KB, containing some of
the concepts and instances used in the evaluation
part. A more detailed Image of depression concepts and instances could be found in the Figure~\ref{fig:Depression_KB}. In this section, different steps for constructing depression-related KB are described.

\begin{itemize}

    \item\textbf{Concepts and Instances of Depression-related KB}
    In this section, different components of depression-related KB, namely concepts~(i.e., `Symptoms', `Risk Factors' and `Supportive Symptoms') and their instances are explained. We also describe how different components are linked to each other. We also represent some examples in each part for more clarifications and ease of interpretation.

    \begin{itemize}
    \item\textbf{`Symptoms'}

    There are
    nine
    main depression symptoms, namely `Sleep pattern change', `Appetite/Weight change ', `Diminished ability to think ', `Loss of Interest ', `Psychomotor Problems ', `Suicidal behaviour ', `Depressed Mood ', `Negative self-evaluation ', `Fatigue '. We consider them as
    nine
    instances linked to the `Symptoms' concept.

    Impaired cognitive functioning, being the sub-instance of psychomotor problems, has several instances such as `decision-making difficulty'. On the other hand, `Indecisiveness' is an instance of the concept of `diminished ability to think'. Since indecisiveness and decision-making difficulty both are the same impaired cognitive functioning, they are connected in the KB by a dotted arrow indicative of the `same as' relationship.
    \item\textbf{`Risk Factors'}

    There are four instances of the `Risk Factors' concept. They are `Demographic features', `Previous Episodes', `Personality Trait', `Major non-mood Disorder'. These instances are placed under the `Risk Factors' concept, in the KB. Some of these instances such as Personality Trait consists of some more sub-instances.

    A personality trait is one of the depression risk factors. `Neuroticism' is the most significant instance of the personality trait concept, consisting of the `negative emotion' and `distress' as its instance and sub-instance, respectively. In the KB an instance~(e.g., negative emotion) is linked to its upper concept~(i.e., the personality trait of neuroticism) or its following instance~(e.g., distress) by a solid arrow. Besides, the relationships between instances of a concept to the other concepts or their instances are shown as dotted lines.
    \item\textbf{`Supportive Symptoms'}

    There are 10 instances of the `Supportive Symptoms' concept. They are `Anxiety', `Self-focused attention', `Abnormal Fear/Phobia', `Tearfulness', `Physical pains',  `Special depressive Phrases', `Irritability', `Obsessive Rumination', `Absolute words', `Depression-related unigrams'. Some of these instances such as Depression-related unigrams include some sub-instances such as `Depression-Treatment' and `Depression-Symptoms'. These sub instances are indicative of special terms related to depression treatment or symptoms, for example.

    in addition, `Anxiety' and `self-focused attention' are two instances of depression supportive symptoms concept. Anxiety is caused by excessive distress. Hence, they are connected to each other by a dotted arrow, indicating a `cause in' relationship. In addition, the feeling of worthlessness/self-criticism could lead to increased self-focused attention. Therefore they are also connected by the dotted arrow.

    \end{itemize}
    \item\textbf{Lexical Sources Linked to the Depression-related KB}
    In this research, we use several lexical sources to be linked to the depression-related KB. As Figure~\ref{fig:Lexical_Database} illustrates, our final lexical database is in the form of an excel file. It is constructed by merging four lexicons, namely LIWC2015 lexicon~(i.e., categories related to depression instances)~\cite{pennebaker2015development}, NRC emotional lexicon~\cite{mohammad2013crowdsourcing}, absolute word lexicon~\cite{al2018absolute}, depression Uni-gram lexicon~\cite{de2013predicting}. Different lexical sources, used in the construction of this research lexical database, are described as follows.
    \begin{itemize}
    \item\textbf{LIWC2015: }
    As explained in Section~\ref{Text_Data}, LIWC2015 is one of the main and golden standard computerised text analysis tools. It is a valuable resource that is widely utilised for measuring positive and negative affects~\cite{de2013major, golder2011diurnal}, as well as extracting various word-based features from textual data~\cite{pennebaker2015development, tausczik2010psychological,pennebaker1999linguistic}. LIWC2015 consists of an English dictionary with 90 linguistic, behavioural, and psychological categories. Each category includes a list of related words. In order to construct the depression-related KB, we used all of the categories, related to the depression instances.
    \item\textbf{NRC Emotion lexicon: }
    It is a broad, high-quality English term-emotion association lexicon focusing on 'joy', `sadness ', `anger ', `fear ', `trust ', `disgust ', `surprise ', `anticipation'~\cite{mohammad2013crowdsourcing}. To enrich the categories word list, we merge them with proper categories of NRC Emotion lexicon~\cite{emotionlexicon}, including  For example, we merge all the words in the `negative emotion' categories of both LIWC and NRC, together. On the other hand, some NRC categories such as `disgust', being an instance of negative emotion in depression-related KB, was separately used to form the list of related words of `disgust' instance.
    \item\textbf{Absolute word lexicon: }
    Depressed individuals mostly talk in absolute language and use greater percentage of absolutist words in their natural language and daily communications~\cite{serani2011living,al2018absolute}. In this study we use the absolute word list, developed by Al-Mosaiwi et al.~\cite{al2018absolute}.
    \begin{landscape}
    \begin{figure}[h]
        \centering
        \includegraphics[scale=0.7]{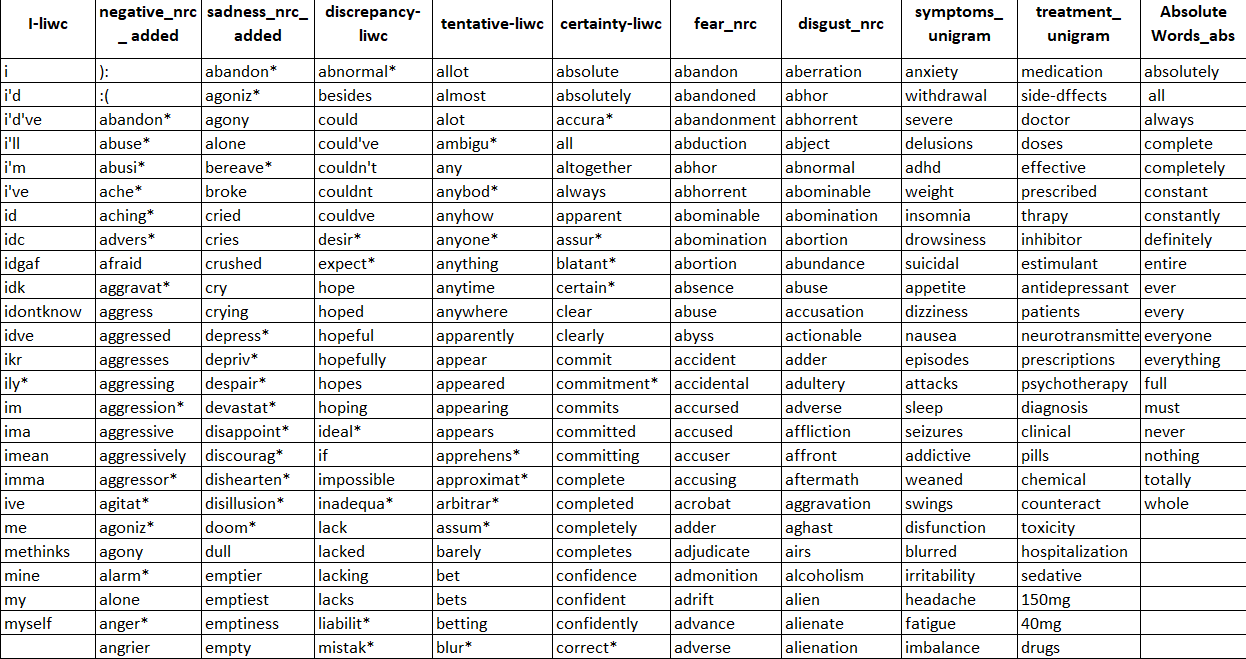}
        \caption{A snapshot from a part of lexical database used in this research. It is constructed by merging four lexicons, namely LIWC2015 lexicon~(i.e., categories related to depression instances)~\cite{pennebaker2015development}, NRC emotional lexicon~\cite{mohammad2013crowdsourcing}, absolute word lexicon~\cite{al2018absolute}, depression Uni-gram lexicon~\cite{de2013predicting}.}
        \label{fig:Lexical_Database}
    \end{figure}
    \end{landscape}
    \item\textbf{Depression Uni-gram Lexicon:}
    Based on a study by Chouldhary et al.~\cite{de2013predicting}, depressed people tend to use special depression unigrams and terms in their communications. Through their study they developed a lexicon of some depression-related categories. In this study, we use those lexicon to make a more comprehensive lexical database.
    \item\textbf{Suicidal phrases source:}
    Based on Fernandes et al., there is a list of terms and phrases, used to detect and extract documents that may include references to attempted suicide~\cite{fernandes2018identifying}. This list may aid in the identification of suicidal thoughts and attempts in textual data\footnote{https://github.com/andreafernandes/NLP\_Tools\_Development}. Hence, in this research, we leveraged this list of words and phrases, aiming
    to calculate
    a score for this symptom. Table~\ref{tab:suicidal_phrases} illustrates different phrases related to suicidal behaviours.
    \begin{table*}[htbp]
        \centering
        \includegraphics[scale=0.4]{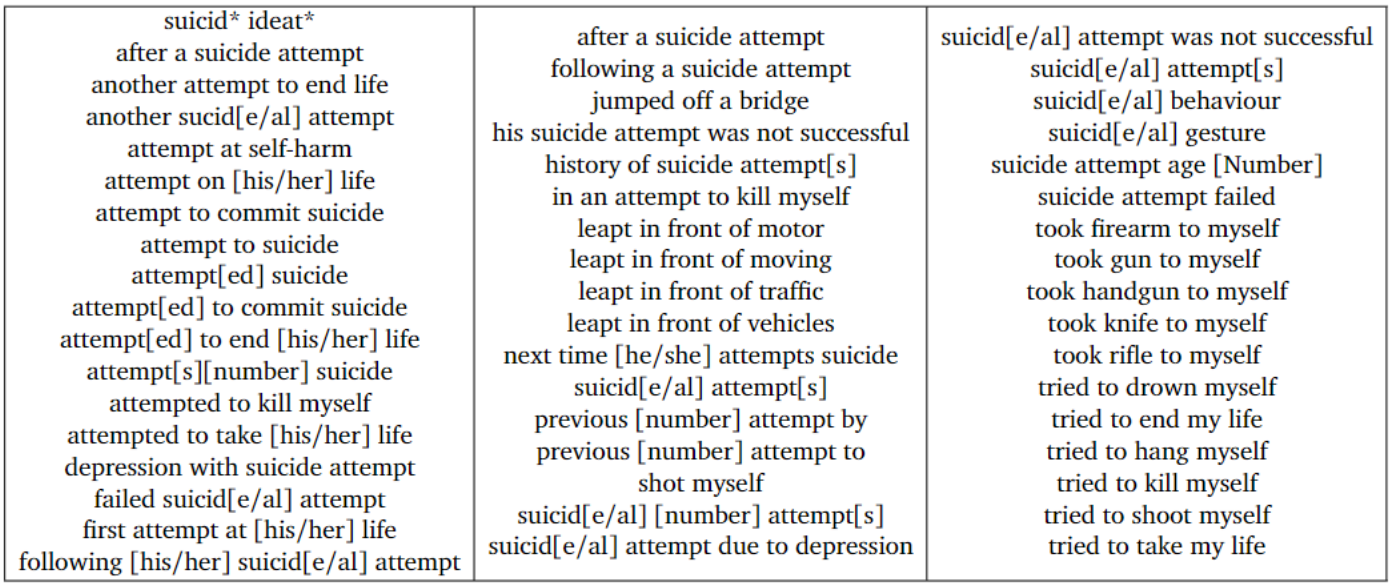}
        \caption{List of terms and phrases used to detect and extract documents that may include references to attempted suicide or ideation in this regard~\cite{fernandes2018identifying}.}
        \label{tab:suicidal_phrases}
    \end{table*}
    \end{itemize}
    \item\textbf{Depression-related KB's APIs: }
    In depression-related KB, we deployed several APIs, aiming at calculating instance-related scores. There are three kinds of APIs used in our KB construction. The task and the structure of each API are explained as follows.

    \begin{itemize}
    \item\textbf{Instance to Lexicon Connector API: }
    In this step, each depression-related instance is empowered by an APIs, which is linking the instance to the proper lexical source. For example, there is a special API that connects the `Absolute word' instance to the `Absolute word lexicon', also another API links the suicidal behaviour instance to the suicidal phrase's source, etc.
    \item\textbf{Cosine Similarity Calculator API: }~\label{cosine-similarity-API}
    In addition to the Instance to Lexicon Connector API, most of the instances are linked to another API, which provides different instances with the cosine similarity scores. This API, which is linked to the instance-related lexical source, uses different NLP techniques~(e.g., TFIDF and Cosine Similarity) to calculate those scores.

    In those APIs that are created for cosine similarity calculation, TFIDF technique is applied, first. For this aim, concatenated texts in each of the 10 chunks for each user~(i.e., mentioned in Section~\ref{data_set}), are considered as different documents. These documents, used in the TFIDF technique, make it possible that important and meaningful words, that are used by each user, to be considered in the analysis. Consequently, we could come up with reliable results.

    \item\textbf{Suicidal Behaviour API: }
    We use a list of phrases, related to suicidal behaviours for identifying the scores of suicidal attempts or ideation from texts. Suicidal behaviour API consists of the regular expressions to count the number of suicidal phrases in texts. the calculated number is used as the calculated score to be used in further stages.
\end{itemize}

\end{itemize}

\subsubsection{Leveraging the Proposed Pipeline to Classify the Textual Data as `Depressed' or `Not depressed'}
\label{depression_classifier}

After constructing the depression-related KB, several calculated instance scores enable us to develop an automated machine learning~(ML) algorithm~(i.e., depression classifier). This algorithm is equipped with the knowledge derived from the mKB and could be used in labelling different texts into `depressed' or `not depressed'. For example, if a text is labelled as `depressed', it shows that the writer of
the corresponding
text was probably depressed. In this approach, we use a data analytics tool for identifying the depression status of a person who wrote a text, instead of receiving direct consultancy or
supervision from experts
or psychologists. Hence, this classifier is considered to be a kind of weakly supervised learning algorithm. This classifier could facilitate many future depression-related studies in terms of creating a large amount of labelled data~(i.e., training data set) for their deep learning analysis. In this section, as shown in Figure\ref{classifier_eval}, we apply the proposed pipeline explained in Section~\ref{Developing-Classifier} to develop a learning algorithm for depression identification by textual data analysis.

\begin{itemize}
    \item\textbf{Pre-Processing and Data Curation: }
    As mentioned before,in Section~\ref{data_set}, our input dataset~(i.e., Reddit posts) contains 820 users' posts~(i.e., posts of each user is
    divided
    in to 10 chunks). They are in the form of `XML' files, containing the textual posts during different time ranges. In the first step, through preprocessing and data curation process, the raw input data should be organised into a proper format. Hence, we use the `XML' and `pandas' packages to turn the XML files into data frame, making the ML analysis feasible. The first data frame consists of 820 rows~(i.e., the number of users) and three columns, namely `user-names', `depression status' and `concatenated-texts'. All the single posts of each user were concatenated together, forming a single long text. This document
    forms
    the content in the `concatenated-texts' column of the data frame. The values in the `depression status' column are `0' or `1', indicative of not-depressed and depressed tags, respectively.

    We also create another data frame~(i.e., the second data frame) containing 8200 rows, each 10 rows related to one of the users. It has the same columns as the previous one. We used the content of this data frame for the `TFIDF' and `Cosine Similarity' scores, mentioned in Section~\ref{cosine-similarity-API}. Afterwards, we use regular expressions to remove stop words, symbols, numbers, and other useless data from the texts. We apply NLP techniques such as the word tokenisation method to the cleaned texts. In addition, all of the words were stemmed using `nltk.stem package'. The output would be lists~(i.e., in the first data frame) or strings~(i.e., in the second data frame) of words, within a specific column in the data frames, which are considered to be our extracted features. A part of Figure~\ref{classifier_eval} is indicative of an example curation process over a text, which is extracted from an XML input file. The final result of curation part, is a list of stemmed forms of extracted and tokenised features.

    \begin{figure}
    \centering
    \includegraphics[scale = 0.55]{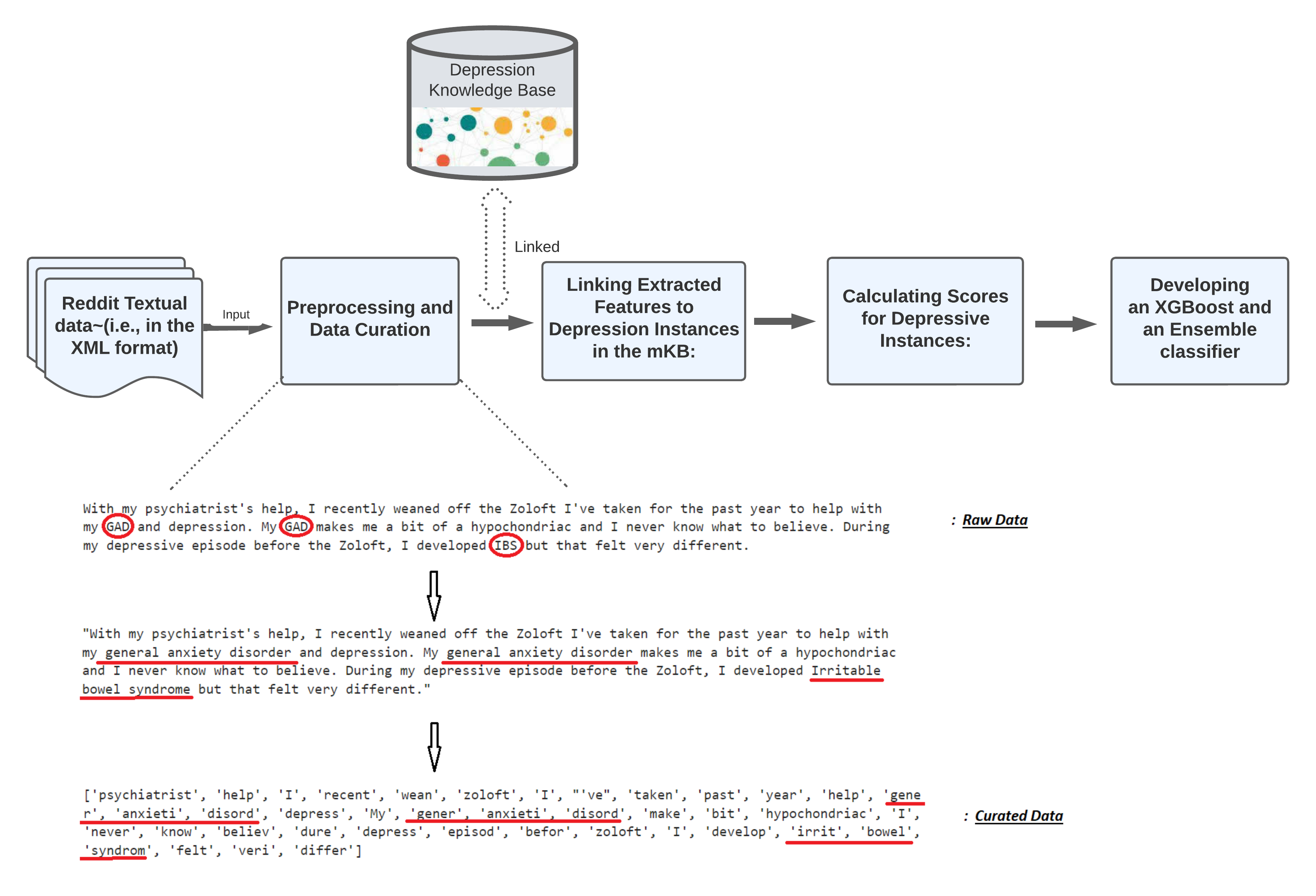}
    \caption{A snapshot of applying proposed pipeline in the Method Section, containing an example curation process result.}
    \label{classifier_eval}
    \end{figure}

    \item\textbf{Linking Extracted Features to the Depression-related Instances in the mKB: }
    For the evaluation part, we use 17 instances of depression KB~(i.e., named in the next part) that could be provided with a proper lexical source. In this part, we assumed every single word that is used in the user's texts is a valuable asset to be considered in the analysis. Hence, we link stemmed and extracted words from input text to the 16~(exclude one out of 17) instances in the mKB. The linked data is provided in the form of a list of stemmed words, related to each user to be used by the corresponding instance-related API for score calculation. The excluded instance is the `Suicidal behaviours' instance. This instance is not linked to the stemmed forms of extracted words, but to the original cleaned text. The aim is to use this original text for calculating a score for the `Suicidal behaviours' instance, which is simply the count of special phrases used in the texts.
    \item\textbf{Calculating Scores for Depressive Instances: }~\label{depression-instances}
    There are 17 depression-related instances in mKB that are provided with score calculator APIs. These instances are related to one of the three depression-related concepts, namely `Symptoms', `Risk factors', and `Supportive Symptoms'. Some instance scores, such as those related to `having physical pain' and `change in appetite' are not extractable by the KB APIs, because the current APIs are enabled by being linked to the lexical sources. Lexical sources are great sources for calculating linguistic- and emotion-related instances, not those related to the individual's physical status.

    \begin{figure}
    \centering
    \includegraphics[scale = 0.67]{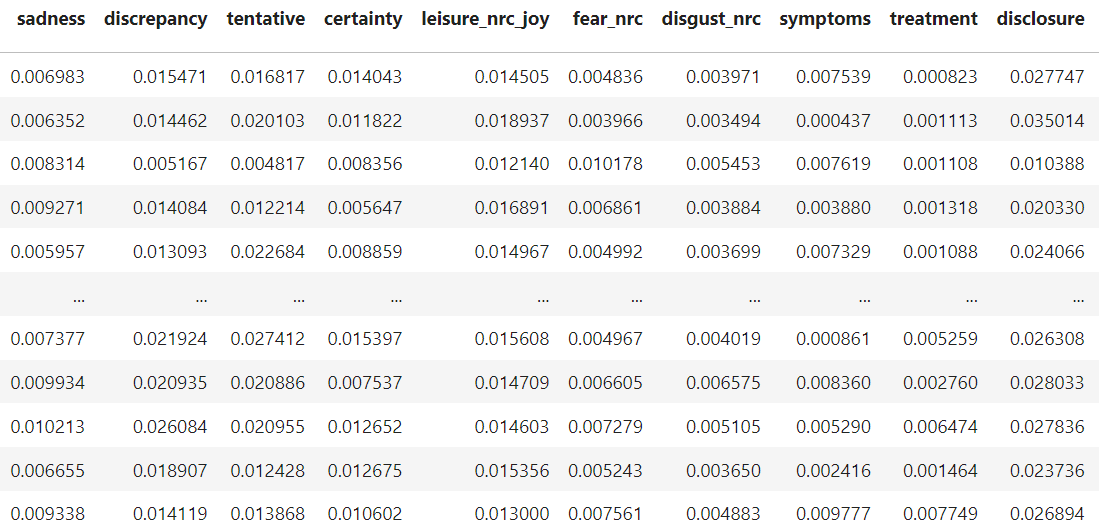}
    \caption{A snapshot of calculated lexical-enabled depressive instance scores.}
    \label{fig:scores-snapshot}
    \end{figure}

    On the other hand, scores related to instances like `change in the sleep pattern' and `social activities' could be calculated by analysing the related metadata. For example, the time and the date of having an activity on social media, the number of followers, or the people who are followed by a user, are useful metadata for calculating some more instance scores. in this research, we are just focusing on calculating the lexical-enabled instance scores, regardless of other metadata-based scores. Consequently, 17 depressive instances are considered to be used in our evaluation. Their related scores~(i.e. for 16 instances) would be calculated and used for developing the depression classifier. Those features, as shown in Figure~\ref{fig:scores-snapshot}, being in the form of calculated cosine similarity scores are fed into the classifier, as its input features) in the next step.

    In addition, the suicidal behaviour scores, calculated by the corresponding API in the KB, are not in the form of cosine similarities. They are integers, ranging between `0' to `31'. On the other hand, calculated cosine similarity scores ranged from `0' to `1'. Hence, we normalised these instance scores, prior to being fed to the classifier as its features. The normalisation process is conducted by applying 'StandardScaler' technique from `sklearn.preprocessing' package, in Python. The depressive instances and their corresponding concepts, which are considered in our evaluation, are as follows:

    \begin{itemize}
        \item\textbf{Concept~1: } Risk factor-related features are `Negative Feeling', `Disgust'
        \item\textbf{Concept~2: } Symptoms-related features are `Sadness', `Discrepancy', `Tentativeness', `Certainty', `Leisure', `Suicidal behaviours'
        \item\textbf{Concept~3: } Supportive symptoms-related features `Self-focused attention', `Anxiety', `Anger',  `Fear',  `Symptom unigrams', `Treatment unigrams', `Disclosure unigrams', `Relationship unigrams', `Absolute words'
    \end{itemize}
    %
        \item\textbf{Classifier: } To identify the best algorithm
        for
        our classifier, we tried several simple and complex models. The implementation results for several simple models such as KNN~(k nearest neighbour), SVM~(Support Vector Machines), Linear Regression, and Naive Bayes were unpromising. The classification reports for all of them were indicative of `0' scores as the outputs for precision, recall and f1-score regarding the depressed class~(i.e., labelled as `1' in the report). Figure~\ref{SVM_REPORT} shows an example classification report for the SVM algorithm. This result derives from the incapability of these models
        to deal
        with our imbalanced data.
        \begin{figure}
            \centering
            \includegraphics[scale = 0.7]{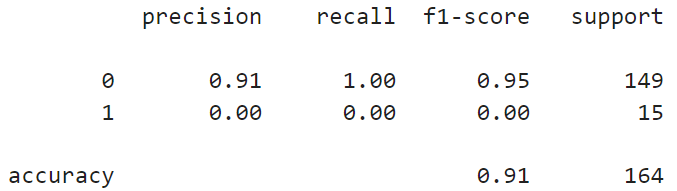}
            \caption{The classification report of the SVM classifier.}
            \label{SVM_REPORT}
        \end{figure}

    Therefore, we decided to implement two ensemble techniques, namely an XGBoost and a stacking model, being more complex and robust models
    compared to
    the previous ones. XGBoost is an ensemble model that enables applying several hyper-parameters to control the learning process~\cite{niu2020ensemble}. This model teaches each of its predictors sequentially based on its predecessor's error. It also makes it possible to give weights to the majority and minority classes, consequently
    eliminating
    the issue of having imbalanced data. Figure~\ref{fig:first-xgboost-result} illustrates the classification report of developed XGBoost classifier with 0.82 accuracy. Applying 10-fold-cross validation, also, leads to an increase in accuracy to 0.87.

    We also developed a stacking model. It contains five stacked classifiers, namely a Random Forest~(RF), a KNN, an SVM, a Naive Bayes, and an XGboost. We removed the imbalanced data issue by down-sampling. To do so, Each of the five stacked classifiers are trained and tested on a separate sample, which are created using 78 depressed users' data and randomly selected one-fifth of not depressed individuals. So we could implement SVM, and other simple models by using the balanced data. We aimed at building a model to analyse our data considering both linear~(e.g., SVM) and non-linear~(e.g., RF and XGboost) approaches in decision making process. Figure~\ref{fig:ensemble_model} is illustrative of the developed ensemble model and the classification report for the test data tagged with `1', showing depressed L.
    \begin{figure}
     \centering
     \includegraphics[scale = 0.5]{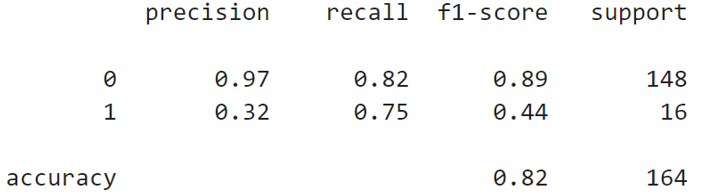}
     \caption{ The Classification report of the developed XGBoost model for depression identification by textual data analysis.}
     \label{fig:first-xgboost-result}
    \end{figure}

     To compare the prediction results of the developed algorithms on a single unseen text, one of the samples in the data set~(i.e., related to a depressed user) were extracted and removed from the data set. Then, it was fed to both of the models. the prediction results for XGBoost and stacking models showed a probability of 0.90 and 0.77 for being depressed, respectively. Taking into account the classification reports, the stacking method could be considered a better method in this regard. Consequently, developed stacked algorithm could be used as a reliable weakly supervised learning model to be used in labelling large data sets~(i.e., `depressed' or `not depressed' tags).
\end{itemize}

\subsection{Evaluation}

%
Through this research, we had two contributions: constructing the depression-related KB, and  developing a depression-related classifier.
To evaluate our proposed approach,
we conducted a survey and requested domain experts to provide their feedback.
%
%
We design three hypopaper to validate the knowledge in the mKB as well as validating the proposed methodology.
To achive this goal, we invited participants with knowledge and backgroung in psychology and coginitiev science, as well as participants with AI and ML skills.
\begin{figure}
\centering
\includegraphics[scale = 0.4]{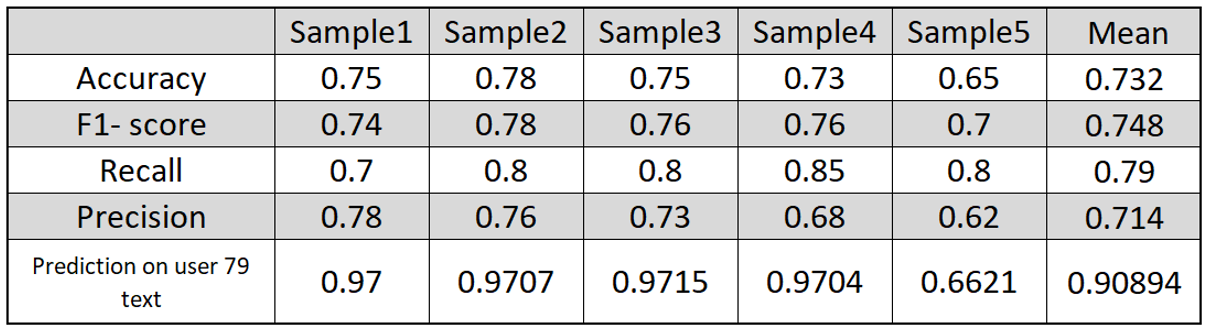}
\caption{The classification report of the developed ensemble model for depression identification by textual data analysis.}
\label{fig:ensemble_model}
\end{figure}
\begin{itemize}
    \item\textbf{Hypopaper 1~(H1):}The construction and components of the depression knowledge base, namely  "Symptoms", "Risk Factors", and "Supportive Symptoms" are relevant to depression Identification.
    \item\textbf{Hypopaper 2~(H2):}Depression-related instances that are used for developing the Machine Learning algorithm are relevant to the identification of depression patterns.
    \item\textbf{Hypopaper 3~(H3):} The structure of the developed depression classifier leads to a reliable results from analysing textual data.
\end{itemize}

\subsubsection{Experiment Setup}

We carried out the experiment in a controlled setting. In our survey, there were two groups of participants, namely psychology-related~(first group) and computer science-related~(second group) participants. The first group
consisted of
seven participants with psychological expertise, consisting of psychology university professors,
Master's and last year
bachelor's students.
The second group consisted of eight participants that were mainly chosen from Ph.D. students with computer science and AI-ML expertise at Data Analytics Research Lab\footnote{https://data-science-group.github.io/}. Through the questionnaire, the participants were provided with some explanations, examples, and instructions to figure out how the questionnaire was organised as well as understanding the aim and the approach of our research. We also arranged a meeting with the participants of the computer science group to present the cognitive and psychological aspects of our research in more depth. The content of
the descriptive
parts of the questionnaire
include the following information:

\begin{itemize}
    \item Brief description of our motivating scenario and the structure of the questionnaire.
    \item Brief description of the depression-related KB construction and its components to make the participants ready for answering the questions related to each section.
    \item Brief description of the process of developing the automated depression classifier as well as demonstrating some snapshots from the developed 
    methodology and sample results.
\end{itemize}

\subsubsection{Questionnaire}

To test the
hypotheses, we created a questionnaire and
shared it with the
the participants. The questionnaire was divided into four sections with multiple-choice questions. The participants were instructed to select one alternative depending on their assessment.  We asked each participant to rank the hypotheses' relevancy based on the `Likert scale system', which utilises a five-point scale to allow the participant to give their opinion~(5: Strongly relevant, 4: Relevant, 3: Neutral, 2: Weakly relevant, 1: irrelevant). After asking the main questions, in the last part, we ask participants to give us their suggestions
for improving
our approach and method.

The initial part of the questionnaire consists of four questions. It was used to gather demographic and background information of the participants, and the next three parts were used to test the H1, H2, and H3 hypotheses. In the second and third parts~(i.e., related to H1 and H2), there are several questions asked
to evaluate
the components and the knowledge included in the depression KB. Hence, for validating the H1 and H2, we focus on the responses of the psychology group.

On the other hand, the fourth section~(i.e., related to H3) has four questions.
These questions
target the participants with AI/ML backgrounds.
The aim is to evaluate the technical aspects, which is developing an ML classifier that could act as a weakly supervised tool for identifying depression patterns from textual data. However, to consider the assessment of the psychology group participants for the technical aspect of our method, the first question of this section, is also
shared with the
psychology group participants. Finally, participants from both groups,
were requested to share feedback,
suggestions, and/or ideas to improve our approach and method.

%
\begin{figure}
    \centering
    \includegraphics[scale = 0.55]{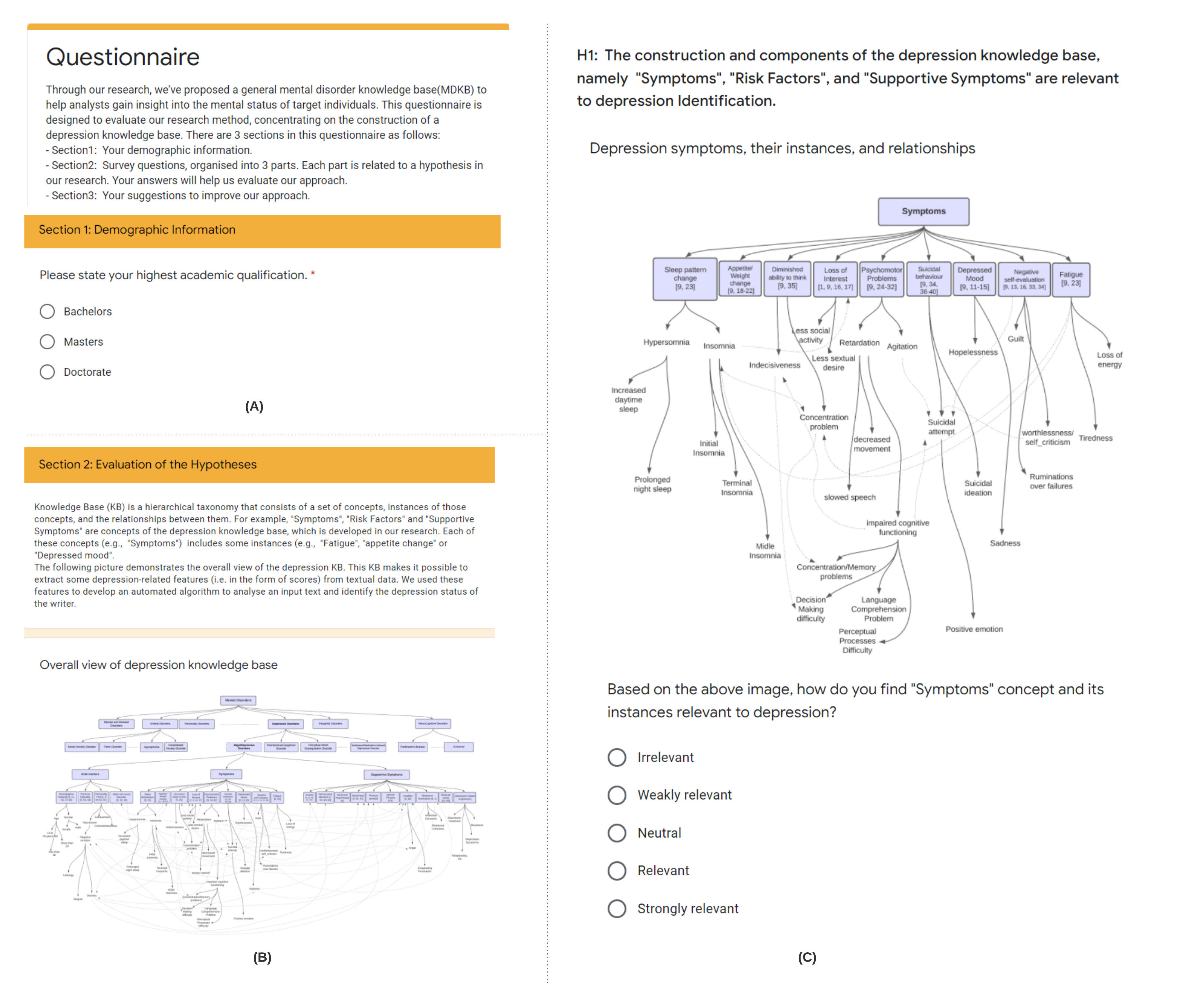}
    \caption{Segments of the questionnaire. (A)~Demographic questions. (B)~An overall view of depression KB. (C)~Evaluation of the first hypothesis associated with`Symptoms' concept and its instances as some components of depression-related KB.}
    \label{fig:questionnaire_1}
\end{figure}
\begin{figure}
    \centering
    \includegraphics[scale = 0.55]{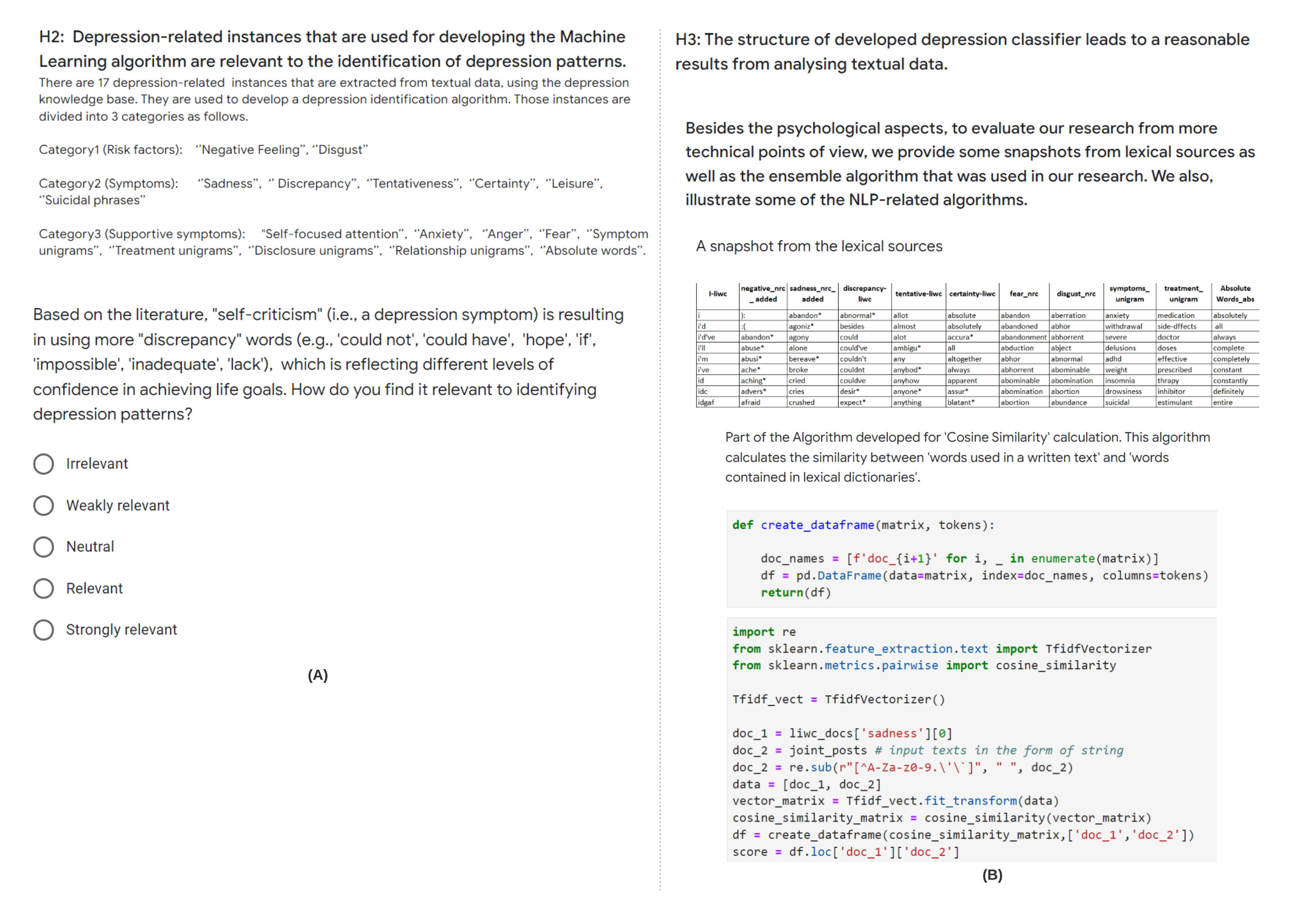}
    \caption{Segments of the questionnaire. (A)~Evaluation of the second hypothesis associated with the depression-related instances that are used for developing the depression identification ML classifier. (B)~Evaluation of the third hypothesis associated with the structure of the developed ML algorithms and depression classifier.}
    \label{fig:questionnaire_2}
\end{figure}
\begin{figure}
    \centering
    \includegraphics[scale = 0.55]{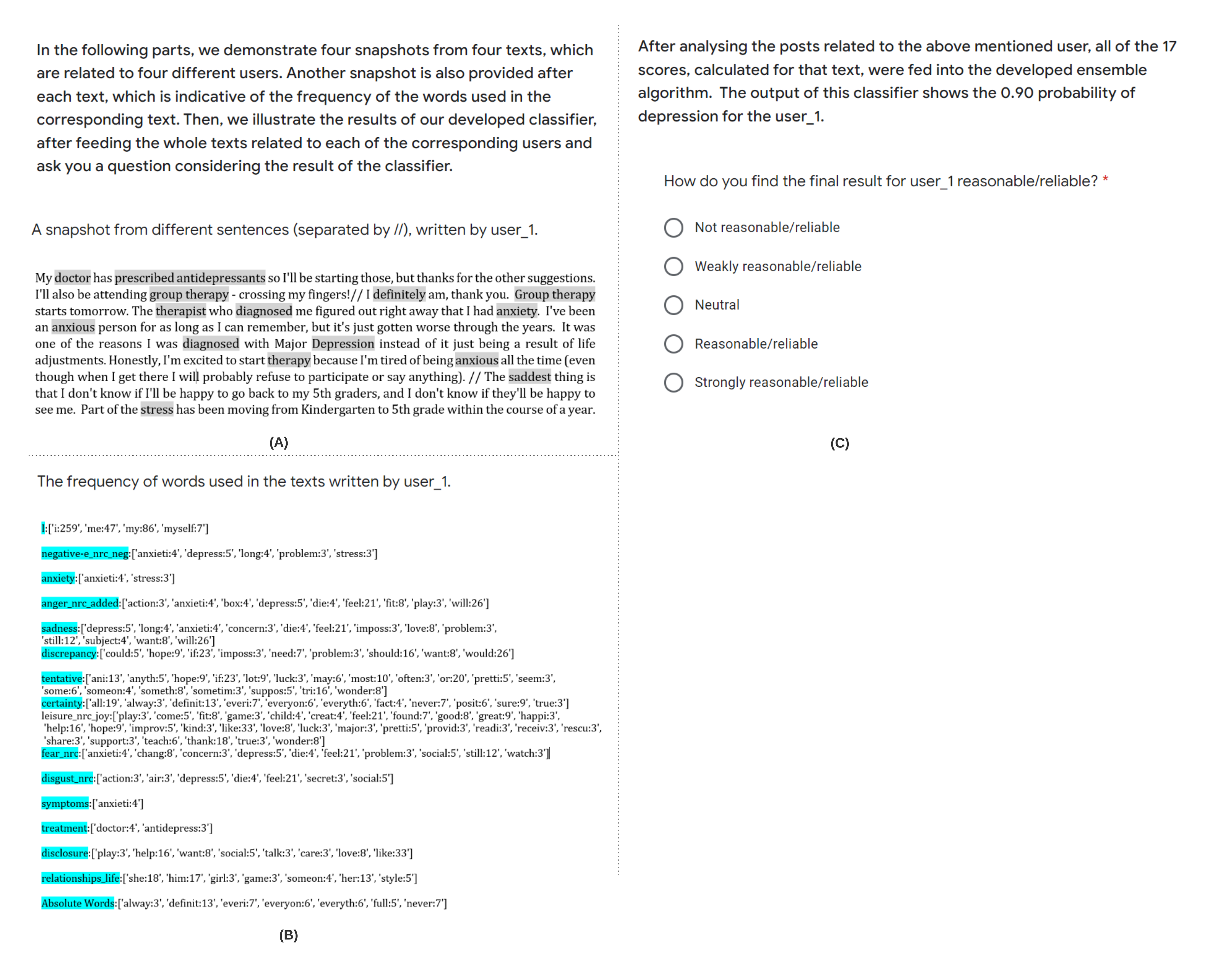}
    \caption{Segments of the questionnaire, related to the evaluation of the third hypothesis.}
    \label{fig:questionnaire_3}
\end{figure}


The questionnaire was created using Google Forms, and a few screenshots from its different sections are shown in Figure~\ref{fig:questionnaire_1},~\ref{fig:questionnaire_2}, and~\ref{fig:questionnaire_3}. Figure~\ref{fig:questionnaire_1}.(A)
relates
to the first part of the questionnaire,
aiming to briefly describe
the motivating
scenario
and the structure of the questionnaire as well as asking for demographic information.
In
Figure~\ref{fig:questionnaire_1}.(B), we briefly presented the definition of `Knowledge base' and also showed the overall view of our depression KB. There are also some snapshots provided in Figure~\ref{fig:questionnaire_1}.(C) and Figure~\ref{fig:questionnaire_2}.(A), related to example questions that are asked from participants for evaluating H1 and H2, respectively. As mentioned before, H3 is designed to evaluate the technical aspects and machine learning approach of our research. Hence, as illustrated in Figure~\ref{fig:questionnaire_2}.(B) we showed some snapshots of the ML algorithms that we developed to help with depression identification. To give the participants a better
view of
the content of the analysed text and the results of these analyses, as Figure~\ref{fig:questionnaire_3}.(A) shows, a short part of the corresponding text is included in the questionnaire. We also showed some of the primary results as well as the final result, derived from analysing textual data by those ML Algorithms in Figure~\ref{fig:questionnaire_3}.(B\&C). 

\subsubsection{Survey Results}

In this section, using the data gathered throughout the experiment, we intend to 
evaluate the
hypotheses.
\begin{figure}
    \centering
    \includegraphics[scale = 0.55]{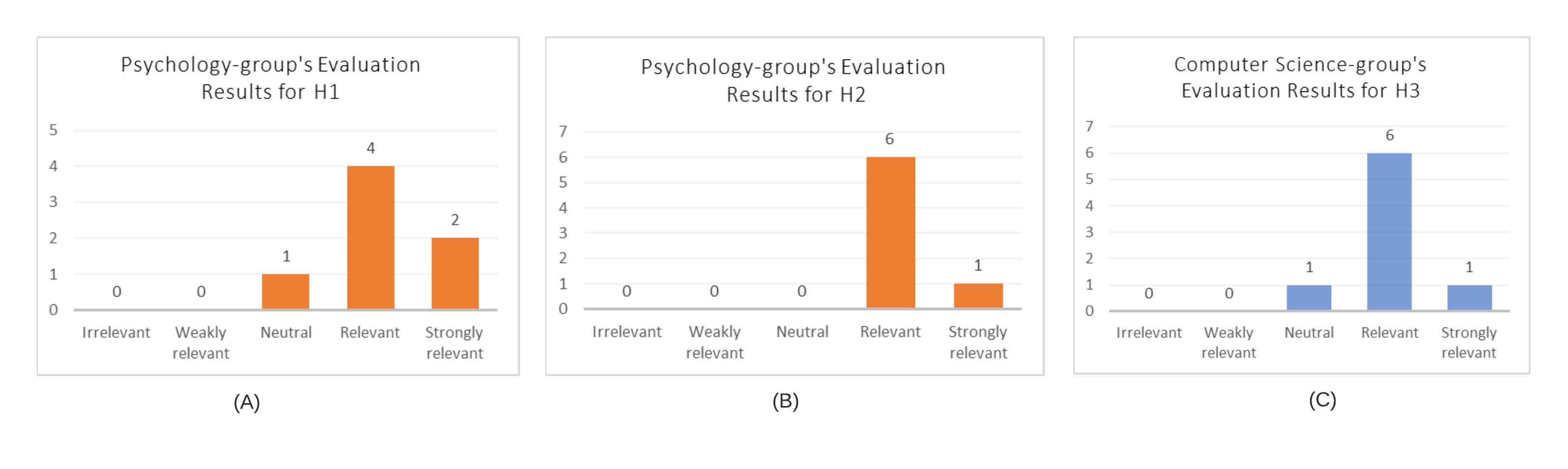}
    \caption{Evaluation of the hypotheses based on data collected during the survey. (A) Evaluation of depression KB construction. (B) Evaluation of instances that are used for developing depression classifier. (C) Evaluation of depression classifier structure.}
    \label{fig:Main_Evaluation}
\end{figure}

\begin{figure}
    \centering
    \includegraphics[scale = 0.55]{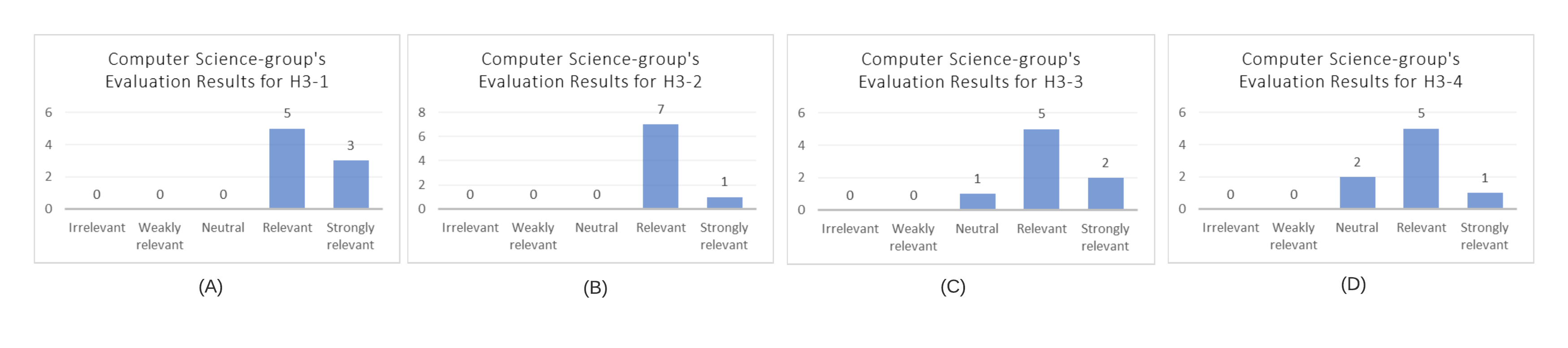}
    \caption{A demonstration of the evaluation results for each of the four questions related to the depression classifier results, that are asked for validating H3.}
    \label{fig:H3_Evaluation}
\end{figure}

\begin{itemize}
    \item\textbf{Evaluation of H1: } Hypopaper 1 presumes that the construction and components of the depression knowledge base are relevant to depression Identification.
    To validate this hypopaper, the Psychology-group answers are considered.
    Figure~\ref{fig:Main_Evaluation}.(A) demonstrates that totally, all of the participants recognised that concepts and instances of developed KB are relevant
    to the identification
    of depression as a mental disorder. Participants
    were asked questions
    related to different concepts of depression KB, namely "Symptoms", "Risk Factors", and "Supportive Symptoms".
    All of
    the participants but one found the components of the developed KB relevant to depression. Figure~\ref{fig:H1_Evaluation} demonstrates the evaluation results for each of the concepts.

\begin{figure}
    \centering
    \includegraphics[scale = 0.55]{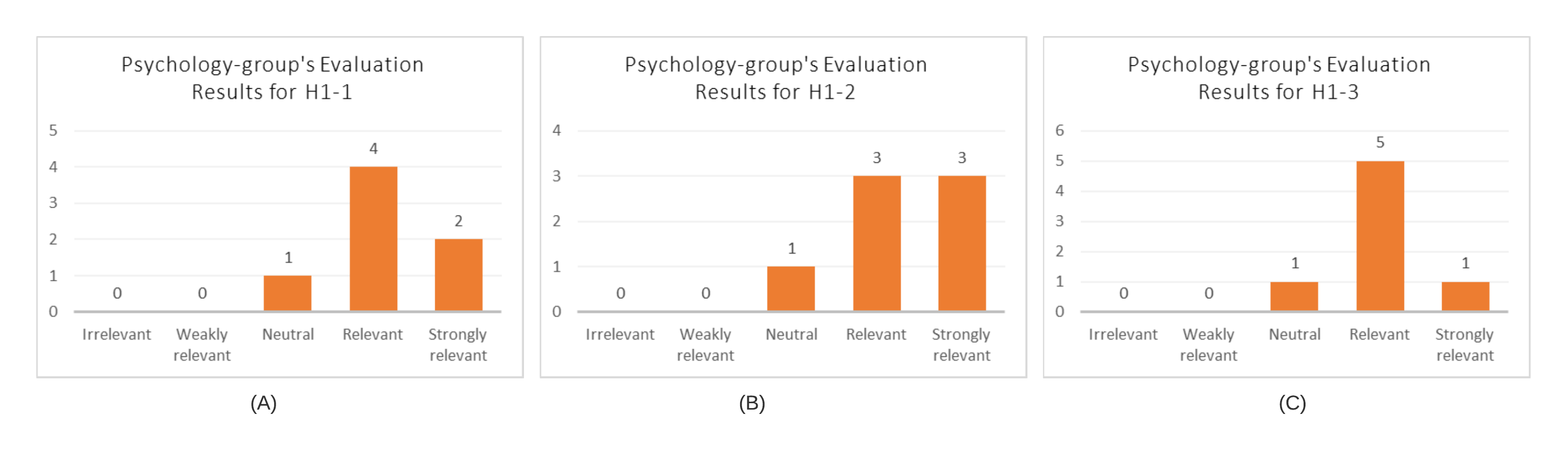}
    \caption{A demonstration of the evaluation results for each of the three questions related to  concepts of depression KB, aiming at validating H1.}
    \label{fig:H1_Evaluation}
\end{figure}
\begin{figure}
    \centering
    \includegraphics[scale = 0.55]{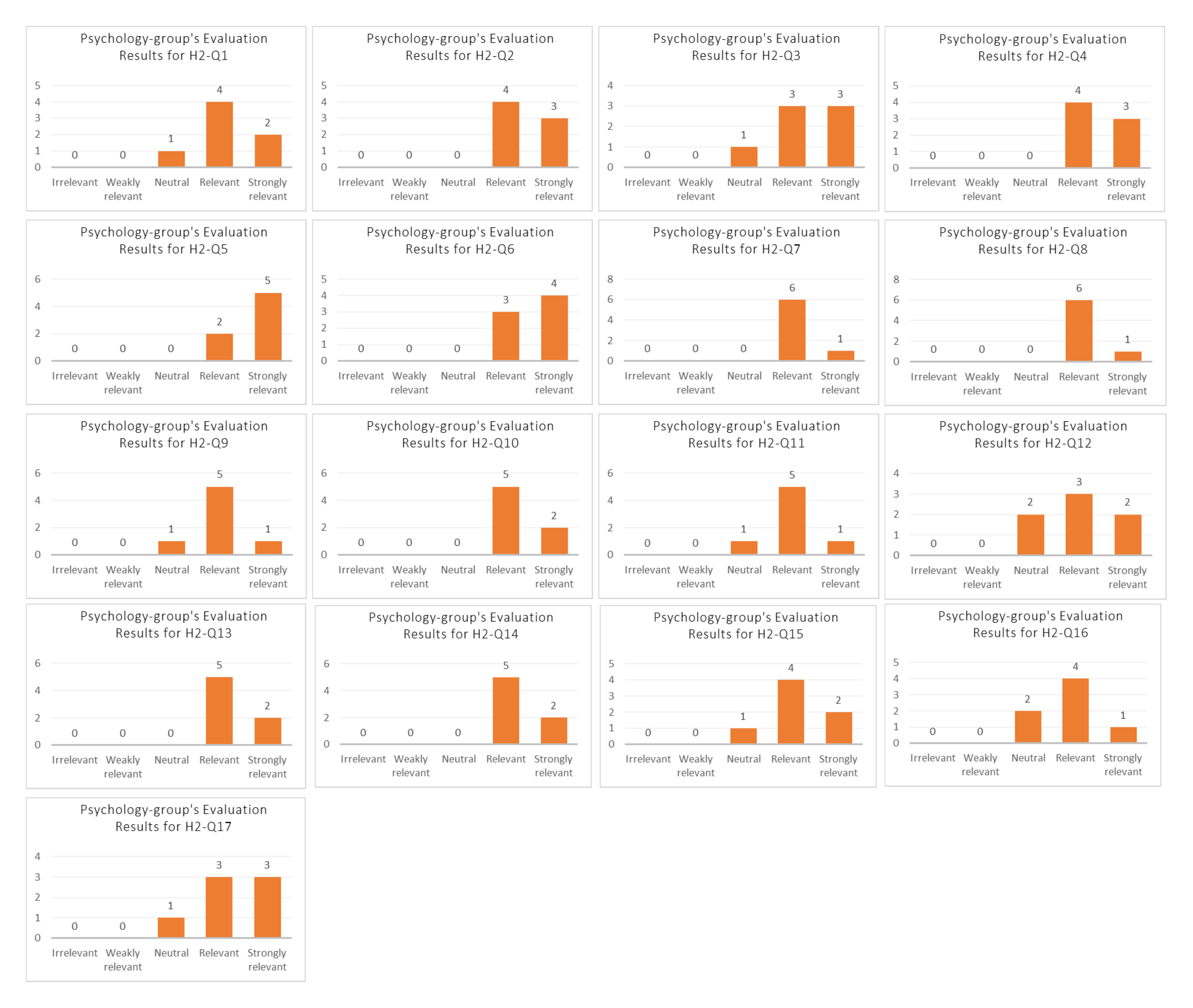}
    \caption{A demonstration of the evaluation results for each of the 17 questions for depression-related instances, leveraged in developing depression classifier, aiming at validating H2.}
    \label{fig:H2_Evaluation}
\end{figure}
    \item\textbf{Evaluation of H2: } Hypopaper 2 presumes that the Depression-related instances
    used for developing the classifier are relevant to the depression identification. To validate this hypopaper,
    the answers from Psychology-group are considered. Figure~\ref{fig:Main_Evaluation}.(B) demonstrates that totally, all of the participants recognised that the instances that are used for implementing
    the developed
    classifier are relevant to
    the identification
    of depression patterns. Participants were
    asked
    several questions related to 17 instances.
    All participants
    found that the instances are relevant 
    to depression
    identification. Figure~\ref{fig:H2_Evaluation} demonstrates the evaluation results for each of the instances.

    \item\textbf{Evaluation of H3: } Hypopaper 3 presumes that the structure of the developed depression classifier leads to
    reliable results
    from analysing textual data. To validate this hypopaper, the
    AL/ML group
    answers are considered. Figure~\ref{fig:Main_Evaluation}.(C) demonstrates that totally, all of the participants recognised that the approach for developed
    ML algorithm is
    relevant to
    the identification of
    depression through textual data analytics. Participants were asked for their assessment regarding the final results of
    the depression
    classifier after analysing four different texts written by potential depressed individuals. All of the participants but one found that the construction of depression classifier leads to
    reliable results
    from analysing textual data. Figure~\ref{fig:H3_Evaluation} demonstrates the evaluation results for each of the four questions.
\end{itemize}

\subsubsection{Discussion}

Since there are some
validity issues,
the results of our study should not be taken as conclusive. Although the overall results of the survey support the H1, H2, and H3, there is still
room for improving the proposed approach.
As Figure~\ref{fig:H2_Evaluation} illustrates, some of the instances used for developing the depression classifier~(i.e., referring to H2) were rated as `neutral' by some of the participants, e.g., instances related to the question three~(i.e., shown as H2-Q3), nine and eleven.

There are seven
instances that were rated by one or two participants as `neutral', namely, `disgust', `self-focused attention', `disclosure', `relationship, life', and `absolute words'. There could be several reasons for such
issues. e.g.,
insufficient clarification and explanation prior to completing the questionnaire. These issues
will be further explored,
in our future works. In addition, based on the comments received from participants and the lessons learned during this study, some future improvements
are considered in Section~\ref{conclusion_future_works}.


\section{Conclusion and Future Work}
\label{conclusion_future_works}
This Section highlights the paper’s contributions and discusses possible future research areas.

\subsection{Conclusion}

Mental health is a key issue around the globe today. Most governments have been engaged in handling the COVID-19 pandemic since it broke out in 2019. As a result of global success in vaccination, the prevalence of the Covid-19 has been considerably reduced. However, for most countries, a major concern now is coping with the consequences of years of viral infection, particularly the psychological and economic consequences. COVID-19 has changed the way we live and work. These shifts can make us feel disappointed and resented, all of which can have a negative impact on our mental health.

As a result, governments
attempt
to assist families and business owners in dealing with the negative results and improving the situation~\cite{beheshti2016business}. Mental problems must be diagnosed before any assistance may be provided. But, Identifying mental disorder symptoms and their patterns, might be difficult due to their intricacy. As a result, it is critical to accurately diagnose mental health concerns and enable their treatment. As a serious mental health problem, depression is one of the top causes of disability around the world. It is a significant contributor to the global illness burden and has the potential to become a serious health problem.

Through this research, we proposed a domain-specific Knowledge Base (KB) to integrate the clinical knowledge regarding symptoms, risk factors, and supportive symptoms that are useful in recognising mental disorders. The KB's knowledge comes from cognitive and psychological studies, and also prior practices in the field. Additionally, in current studies, hand-labelled training sets are used to identify mental disorder patterns from textual data, especially when a domain expert's knowledge is needed. This task could take a long time and be costly. We provide a weaker form of supervision by enabling the generation of training data from developed KB.

\subsection{Future Works}

Considering the comments from participants in the preceding Section, as well as the lessons learned from this study, this research's approach could be improved from two different points of view. The first is improving the construction of the developed mental disorder Knowledge base~(mKB), and the second is bettering the technical aspects of the developed mental disorder classifiers. In the following sections, these two viewpoints are described more.

\subsubsection{Components of Mental Disorder Knowledge Base}

Based on the results derived from the survey, conducted in Section~\ref{Evaluation}, we found that there are some instances in the KB that got rated as `neutral' by some of the participants, e.g., `disgust', `self-focused attention', `disclosure', `relationship, life', and `absolute words'. To improve the validity and relevancy of components of developed KB to depression identification patterns, in our future work, we will consult and interview some experts who have a high academic background~(i.e., at least with PhD degree) or more than five years of experience as a therapist to see their opinion and get their comments in this regard.

Hence, we would improve the construction of the KB to a user-friendly and interactive mode, aiming at enabling the `add item' or `remove item' capabilities to the KB. For example, if after consulting with therapists, it becomes clear that the `disgust' instance is not relevant enough to the depression identification patterns, we could easily remove it from our KB and consequently from the corresponding classifiers. On the other hand, if during our interview sessions it turns out that other instances could be added to the KB, It will quickly be done by using the `add item' capability of the KB.

In addition, the instance-related APIs of the KB, were mainly constructed based on the textual analysis over single words~(i.e., unigrams), that are included in a text.
As future work,
we would consider analysing the bi-grams~(i.e., a pair of consecutive words) and tri-grams~(i.e., a group of three consecutive words) to have
a more
effective text analysis. Therefore, The APIs could recognise the sequential meaning of the words. For example, the difference between `am not happy' and `am happy' could be detected using these APIs.

\subsubsection{Technical and Classifier development Processes}

As one of our future works, we would leverage more sophisticated learning algorithms such as deep neural networks and rule-based approaches, to consider more complex mental disorder patterns in our analysis~\cite{khatami2020convolutional,tabebordbar2018adaptive}. For example, `Convolutional Neural Networks' could enable the classifier to mine the hidden and complex patterns of mental issues that lay behind the input data. Also, using rule-based approaches could help us in developing a mental
health-related
knowledge graph, including an inference engine, aiming at gaining more accurate insights~\cite{batarfi2015large,tabebordbar2020conceptmap}.
These approaches could help us improve the reliability of the results derived by our weakly-supervised classifier for labelling training data sets.

Mental disorders~(e.g., depression) are mainly developed over time. On the other hand, learning algorithms such as time series are learning models to forecast future outcomes based on prior observed values. Hence,
it could be
an ideal approach for predicting the mental status of potential individuals suffering from mental disorders. Therefore, we would consider this approach in our future studies.


\section*{Acknowledgements}
- I acknowledge the AI-enabled Processes (AIP\footnote{https://aip-research-center.github.io/}) Research Centre for funding My Master by Research project.

\bibliographystyle{abbrv}
\bibliography{ms}

\end{document}